\documentclass[twocolumn,preprintnumbers,showpacs,amsmath,amssymb,prl]{revtex4-1}
\usepackage{graphicx}
\usepackage{amsmath, amsthm, amssymb}
\usepackage{latexsym}
\usepackage{amssymb}
\usepackage{bm}
\usepackage{dcolumn}
\usepackage{epstopdf}

\usepackage{color}

\newcommand{\et}{{\it et al.}}

\newcommand{\rt}{$T_1^{-1}$}
\newcommand{\rtt}{$(T_1T)^{-1}$}
\newcommand{\rttwo}{$T_2^{-1}$}
\newcommand{\Ox}{$^{17}$O}
\newcommand{\YBCO}{YBa$_2$Cu$_3$O$_{7-x}$}
\newcommand{\BSCCO}{Bi$_2$SrCa$_2$Cu$_2$O$_{8+\delta}$}
\newcommand{\TBCO}{Tl$_2$Ba$_2$CuO$_{6+\delta}$}

\begin{document}

\title{Nuclear Magnetic Resonance Studies of Vortices in High Temperature Superconductors}
\author{A.M. Mounce, S. Oh, and W.P. Halperin}
\affiliation{Department of Physics and Astronomy, Northwestern  University, Evanston, IL 60208, USA}

\date{Version \today}


\begin{abstract}\vspace{.25cm}
 The distinct distribution of local magnetic fields due to superconducting vortices can be detected with nuclear magnetic resonance (NMR)  and used to investigate   vortices and related physical properties of extreme type II superconductivity.  This review summarizes work on  high temperature superconductors (HTS) including cuprates and pnictide materials.  Recent experimental results are presented which reveal the nature of vortex matter and novel electronic states.  For example, the NMR spectrum  has been found to provide a sharp indication of the vortex melting transition.   In the vortex solid  a frequency dependent spin-lattice relaxation has been reported in cuprates, including \YBCO, \BSCCO, and \TBCO.  These results have initiated a new spectroscopy via Doppler shifted nodal quasiparticles for the investigation of vortices.  At very high magnetic fields this approach is a promising method for the study of vortex core excitations.  These measurements have been used to quantify an induced spin density wave near the vortex cores in Bi$_2$SrCa$_2$Cu$_2$O$_{8+\delta}$.  Although the cuprates have a different superconducting order parameter than the iron arsenide superconductors there are, nonetheless, some striking similarities between them regarding vortex dynamics and frequency dependent relaxation.

\end{abstract}

\maketitle

\section{Introduction}

Among the many tools used by experimenters to explore the properties of superconductors NMR has played an important role.  In this brief review we focus on applications of nuclear magnetic resonance (NMR) to study the mixed state in type II superconductivity.  The interest in this topic and the corresponding explosion in the number of publications where NMR has been of central importance, came immediately after the discovery of high temperature superconductivity\cite{bed86} with  the earliest  NMR and nuclear quadrupole resonance (NQR) papers reporting on (La$_{1-x}$Sr$_x$)$_2$CuO$_4$,\cite{lee87} \YBCO,\cite{tak89} and Tl$_2$Ba$_2$Ca$_2$Cu$_3$O$_{10+\delta}.$\cite{lee89}  NMR is  a microscopic probe of the electronic state and has the advantage of being spectroscopically selective for specific crystallographic positions  in the superconducting compound.  The spectrum is the distribution of local magnetic fields for which there are several contributions from superconductivity.  There are fields from diamagnetic currents circulating at the surface of the sample, especially important at low applied magnetic fields, superposed on the inhomogeneous distributions of magnetic field from vortex supercurrents from the sample interior.   An ideal distribution, calculated from GL theory using algorithms from Brandt,\cite{bra91,mitthesis} is shown in Fig.~1 and 2, called the Redfield pattern.  Additionally,  the spin-lattice relaxation rate, \rt, gives a measure of the electronic excitations which provide important signatures of the superconducting state.  These aspects have been described in the early literature, notably in the review by MacLaughlin\cite{mac76} that emphasizes the importance of NMR in providing early support for the BCS theory.  

\begin{figure}[t]
\centering
	\includegraphics[width=.5\textwidth]{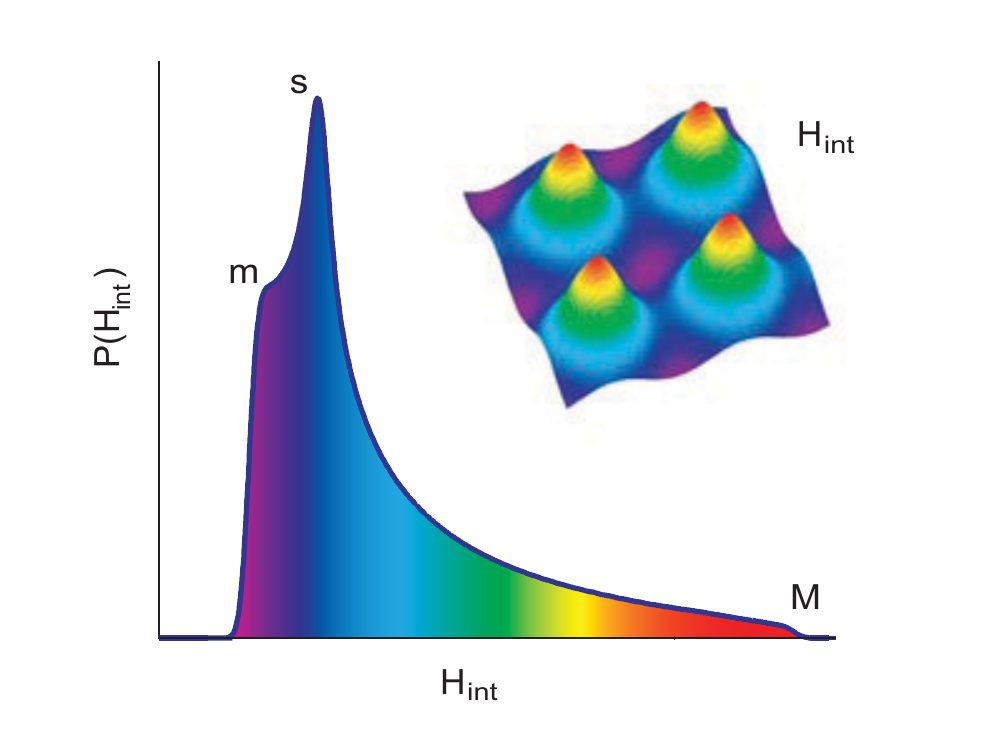}
	\caption{The vortex distribution of magnetic fields equivalent to the NMR spectrum often called the Redfield pattern.  The singularities in the distribution are labeled: the minimum field, m; the saddle point field, s; and the maximum field, M,  at the vortex core. The inset shows the corresponding spatial distribution.  This figure was taken from Mitrovi{\'c} \et\cite{mit01b} calculated from Brandt's algorithm\cite{bra91} for $H = 37$ T.} 
	\label{colorvortex}
\end{figure}
After the discovery of high temperature superconductivity (HTS), a large number of excellent reviews of NMR have been written; among these are: Pennington and Slichter,\cite{pen90} Asayama \et, \cite{asa96} Berthier \et, \cite{ber96} Rigamonti \et, \cite{rig98} Walstedt,\cite{wal08} and Curro.\cite{cur09} This work has been followed by the discovery of new compounds, or new work on previously known materials, that have challenged our understanding of the symmetry of the order parameter in the superconducting state, notably for UPt$_3$, Sr$_2$RuO$_4$, CeCoIn$_5$, MgB$_2$, cuprates, and the iron pnictides and their related compounds.   In this review we will consider aspects of vortex structures in a selection of these materials that have been recently explored using NMR. Two important advances in the past decade in applications of NMR to the understanding of vortex structures in HTS are that high quality single crystals have become available and measurements at very high magnetic fields have become routine.  Most of the focus in this review will be directed toward these developments.  Contributions to our survey are largely taken from the condensed matter NMR group at the National High Magnetic Field Laboratory (NHMFL) in Tallahassee, Florida, where many of these developments were made, and their collaborations with the NMR group at Northwestern University.

\begin{figure}[b]
\centering
	\includegraphics[width= .5\textwidth]{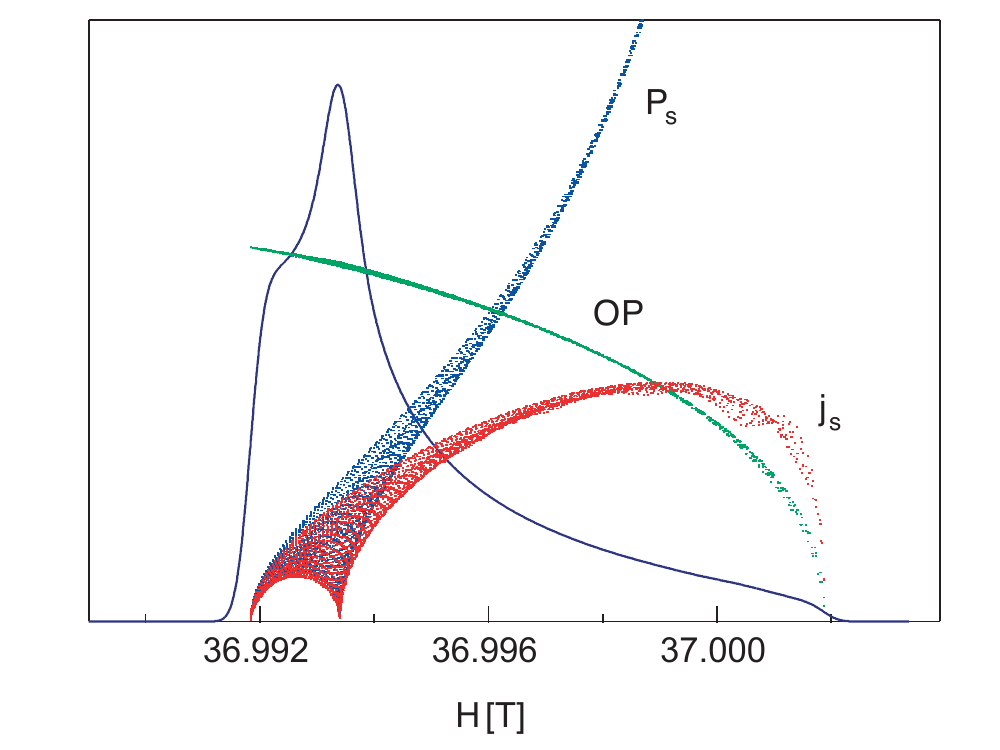}
	\caption{The vortex spectrum, supercurrent momentum, $p_s$, supercurrent density, $j_s$, and order parameter, OP, calculated by Mitrovi{\'c}\cite{mitthesis} for \YBCO\, using Brandt's method\cite{bra91} at $H = 37$ T.} 
	\label{specandps}
\end{figure}

\section{Vortex Spectrum}

It was shown by Abrikosov\cite{abr57} for type II superconductors, $\kappa \equiv \lambda/\xi >> 1$, that quantized vortices penetrate a superconductor with a penetration depth, $\lambda$, and coherence length $\xi$.  The Ginzburg-Landau parameter, $\kappa \sim 100$, satisfies this condition for  HTS.  From the London equations,\cite{lon50} it can be shown that the magnetic field distribution takes the form shown in Fig.~1 and 2 where the quantized flux bundles, or fluxons, are assumed to be rectilinear.  In this London model for the vortex structure the vortex core region is excluded since it is assumed that the coherence length is sufficiently small and the external magnetic field, $H$, is substantially lower than the upper critical field $H_{c2}$.  From Ginzburg-Landau theory\cite{abr57,bra91} it was shown that the vortices form a lattice and within this structure there is a spatially inhomogeneous distribution of magnetic fields, supercurrent momenta, supercurrent density, and the order parameter. Using algorithms developed by Brandt\cite{bra91} these distributions can be calculated for any specific values of $\lambda$ and $\xi$, as is shown in Fig.~2 taken from Mitrovi{\'c}'s calculation\cite{mitthesis}  for \YBCO.  The majority of type II superconductors are highly anisotropic with an anisotropy axis parallel to the crystal $c$-axis.  The flux distribution for $H||c$ can be expected to form a two-dimensional lattice with a high degree of symmetry, such as square or hexagonal, or simple distortions of these.    Such details will alter the Redfield pattern, but not sufficiently to be of concern here.  In fact the properties of the inhomogeneous mixed state are less dependent on this symmetry approaching the vortex core and at high magnetic fields. 

The classic Redfield pattern from the GL-theory is not generally found to be in very good quantitative agreement with NMR experiments.  Usually there are broadening mechanisms in addition to the local fields from supercurrents which are convolved with the Redfield pattern, as for example, in the works on \YBCO,\cite{mit01b,mit03b} Tl$_2$Sr$_2$BaCu$_2$O$_{6.8}$,\cite{zhe02} \TBCO,\cite{kak03}  and \BSCCO.\cite{mou11b}  An example where good agreement was obtained is for the superconducting state of CeCoIn$_5$ shown in Fig.~3.  Another case is the lineshape simulation of the powder pattern of the strongly anisotropic superconductor, MgB$_2$.\cite{che06}  In this instance the frequency shift from the magnetic moment of the screening currents was explicitly included in the GL-calculation along with the corresponding magnetic shift distribution from the vortex supercurrents.  An excellent representation of the full, quadrupolar-split, spectrum was obtained.  In these instances the penetration depth, and with less accuracy the coherence length,  can be inferred as fitting parameters.  However, without a consistent representation of the lineshape from the theory a determination of the penetration depth from either the second moment of the spectrum or the full-width-at-half-maximum is not very meaningful.\cite{son95}

\begin{figure}[t]
\centering
	\includegraphics{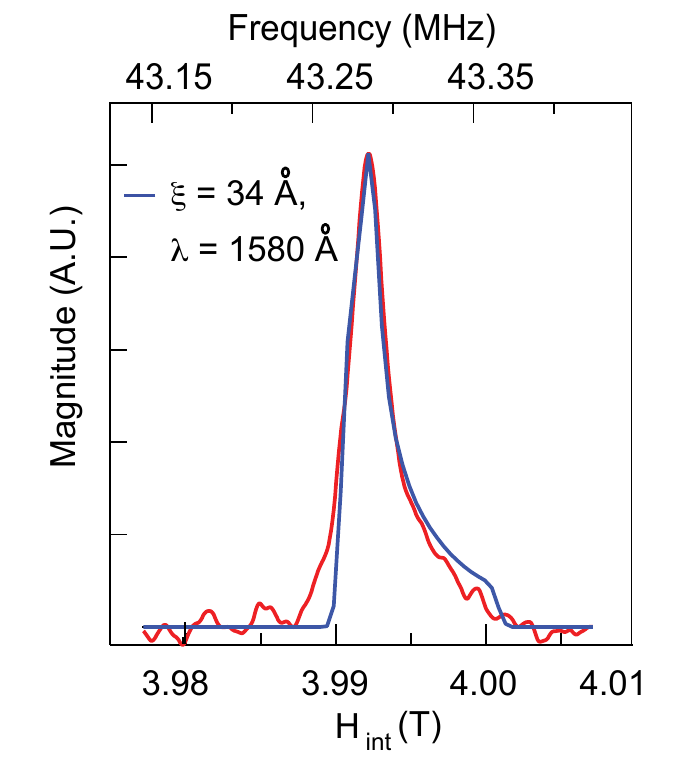}
	\caption{The vortex distribution of magnetic fields in CeCoIn$_5$ taken from Koutroulakis \et\cite{kou08} measured from their $^{115}$In NMR spectrum and compared with their calculated spectrum  from Brandt's algorithm\cite{bra91} at $H = 4$ T.} 
	\label{CeCoIn}
\end{figure}

\section{Vortex phases}

\begin{figure}[t]
\centering
	\includegraphics[width=.45\textwidth]{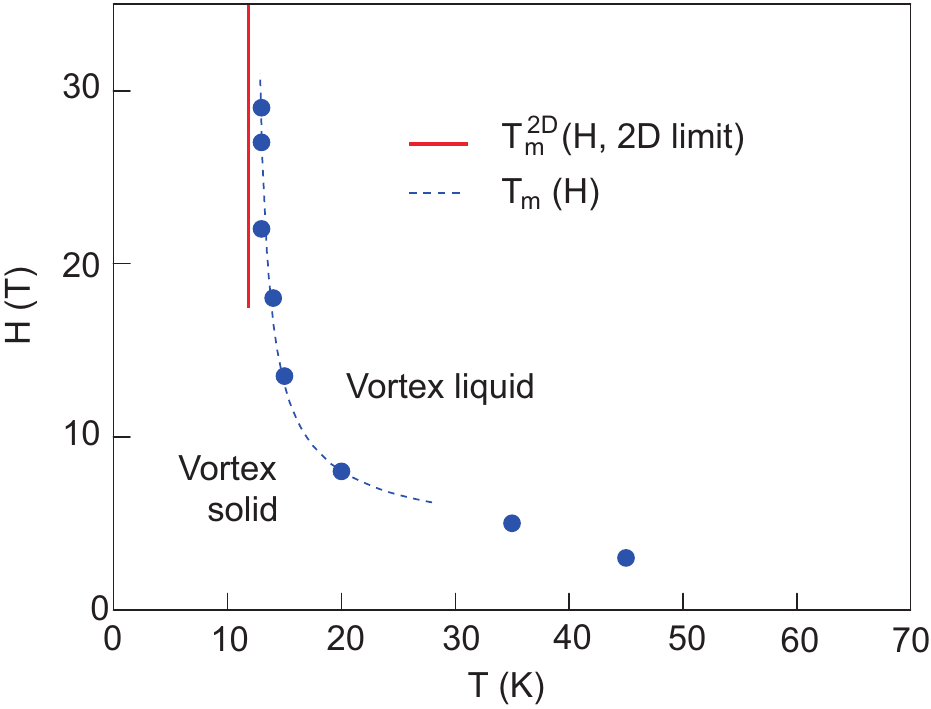}
	\caption{The  vortex phase diagram for \BSCCO\, taken from Chen \et\cite{che07} According to the theory of Glazman and Koshelev,\cite{gla91} there is a limiting value of the melting transition at high magnetic field (solid red line) which, theoretically, should be an ideal two-dimensional transition. From experiment it was found that $T_{2D}=12$ K for an overdoped crystal with $T_c = 75$ K and corresponds to a penetration depth of $\lambda_{ab} = 220$ nm.  The dotted line is a fit to the theory.} 
	\label{phasediagram}
\end{figure}

For NMR to probe the vortex solid structure, vortices must be stationary on the time scale of the NMR experiment.  As thermal fluctuations increase at higher temperatures, vortices enter a liquid-like phase averaging out the inhomogeneous magnetic fields and motionally narrowing the NMR spectrum. For low transition temperature superconductors, the thermal fluctuations are less energetic resulting in a robust solid vortex phase, stable up to $T_c$.  However, in the case of HTS, with its  high transition temperatures and high anisotropy, the result is a stable vortex liquid  that solidifies only at sufficiently low temperatures. In some cases, such as highly anisotropic  \BSCCO,  the liquid phase can extend over a significant region of the magnetic field-temperature phase diagram, Fig.~\ref{phasediagram}. Vortex melting is in fact the only true thermodynamic transition for a clean type II superconductor in a magnetic field.\cite{bla94}  The first order signature of this transition has been extensively studied for both \YBCO\cite{kwo92} and \BSCCO\cite{fuc98,che07}. It is essential that NMR experiments intended to explore spatial configurations of vortices be performed well into the vortex solid domain.  

Formation of a vortex solid has been identified from NMR measuring either spin-spin relaxation, \rttwo, or the NMR linewidth as a function of temperature.  For measurements on \YBCO\, an abrupt change in \Ox\,  \rttwo\, was observed\cite{bac98} and related to a change in vortex dynamics. This corresponded well to the irreversibility line at least at low magnetic fields.  Similar behavior, even more pronounced, has been observed in $^{75}$As \rttwo\, experiments on the electron doped pnictide, Ba(Fe$_{0.93}$Co$_{0.07}$)$_2$As$_2$.\cite{oh11a}  
Additionally, changes in the linewidth have been helpful to identify vortex freezing as reported for \YBCO\,\cite{rey97,bac98} and Ba(Fe$_{0.93}$Co$_{0.07}$)$_2$As$_2$.\cite{oh11a}

In the case of \Ox\ NMR on  \BSCCO\ there is an abrupt signature of the formation of the solid vortex phase at a well defined temperature where the magnetic field distribution increases abruptly on cooling, determined from the second moment of the NMR spectrum.\cite{che07}  In Fig.~4 we show the measurement of the phase diagram which results from plotting this signature, compared to two-dimensional melting theory\cite{gla91} with which there is good agreement at high magnetic field.  The theory shows that the vortex-vortex interactions are largely determined by the electromagnetic coupling between supercurrents;  the  contributions from Josephson tunneling between cuprate layers are relatively suppressed in the limit of high field leading to quasi two-dimensional behavior.  The phase boundary was calculated to approach a vertical asymptote for the limit of an ideal two-dimensional melting transition which corresponds well to  experimental observations.\cite{che07} 

\section{Spatially Resolved NMR}

As it became more evident that the high $T_c$ superconducting cuprates have an order parameter with $d$-wave symmetry, it was recognized that the existence of nodes in the gap at the Fermi surface would affect the thermodynamics at low temperatures owing to vortex supercurrents.\cite{yip92}  The spatially averaged density of states near the Fermi surface was calculated for $d$-wave pairing symmetry\cite{vol93} including  Doppler contributions to the quasiparticle excitation spectrum from the vortex supercurrents.  This term in the energy has the form, $ \delta \epsilon \sim {\bf v}_F$$\cdot$${\bf p}_s$, where ${\bf p}_s\, \propto \,1/r$, and ${\bf v}_F$ is the Fermi velocity, ${\bf p}_s$ is the supercurrent momentum, and $r$ is the distance from the vortex core.  It was  found that the spatially averaged density of states has a unique $\sqrt{H}$ dependence, which became known as the Volovik effect. Bulk experimental probes of the density of states including specific heat\cite{mol94, mol97} and thermal transport\cite{aub99,chi00}  were  in agreement with this picture.  

\begin{figure}[t]
\centering
	\includegraphics{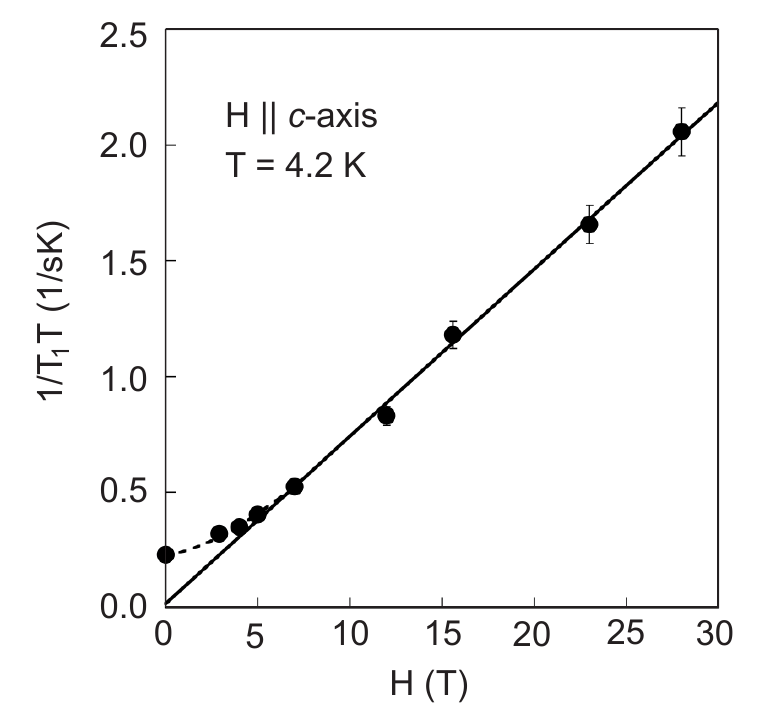}
	\caption{The magnetic field dependence of the $^{63}$Cu NMR relaxation rate is shown for Tl$_2$Sr$_2$BaCu$_2$O$_{6.8}$ taken from Zheng \et \cite{zhe02}.The linear behavior is expected for the average rate according to the Volovik effect for a $d$-wave superconductor.  There is no evidence here for a Zeeman contribution to the quasiparticle excitations which is expected\cite{mit01b} to vary with magnetic field according to $\propto$ $H^{2}$.} 
	\label{zheng}
\end{figure}

Similarly, the average of \rt\, should be proportional to the average of the joint density of states that is expected  to depend on magnetic field,  $\propto H$.  This behavior has been reported in Tl$_2$Sr$_2$BaCu$_2$O$_{6.8}$ by  Zheng \et \cite{zhe02} and is displayed in Fig. \ref{zheng}.  Further discussion of the Volovik effect is deferred to appendix A.  

However, the behavior of the vortex core and extended quasiparticle states on a local microscopic scale were still not well-understood and the need for a local probe of the electronic excitations was evident.  Scanning tunneling microscopy and scanning tunneling spectroscopy are sensitive to the local density of states; but, these probes are limited to the surface of the sample in contrast to NMR.  The sample volume to which NMR responds  is within a London penetration depth, $\lambda$, of the surface, generally a few hundred nm in size.  Superconductivity is suppressed very close to the surface of the sample,  on the much smaller length scale given by the superconducting coherence length, $\xi$, typically a few nm for extreme type II materials.  This leaves a region of the superconductor to which NMR is sensitive, near the surface but sufficiently far away that it is unperturbed by surface effects.

It was predicted theoretically that NMR relaxation would be sensitive to spatially dependent quasiparticle states,\cite{tak99} since the spin-lattice relaxation rate, \rt, depends on the density of states, $N(\epsilon)$.  As pointed out above, in the mixed state of the superconductor, $N(\epsilon)$ must vary in space in response to Doppler contributions from the vortex supercurrents.  With a well-defined vortex lattice spectrum, the local magnetic fields correspond to points in real space, see the inset to Fig.~1. So it was suggested that examining the relaxation at different internal magnetic fields, resolved within the NMR spectrum at different freqiencies, would allow a determination of quasiparticle excitations to be mapped throughout real space in what we call spatially resolved NMR.\cite{mit01b}   In the following we will discuss various examples of spatially resolved NMR experiments on HTS. 
  
\subsection{Spin-Lattice Relaxation: Experiment}

The NMR relaxation rate in a normal metal is calculated in terms of the thermal average of the joint density of states:
\begin{equation}
1/T_1 \propto  \int N(\epsilon_i)N(\epsilon_f) f(\epsilon_i)[1-f(\epsilon_f)]\, \mathrm{d} \epsilon
\end{equation}
where $i$ and $f$ indices label the initial and final electron states which are necessarily spin-up and spin-down  in order to cause a nuclear spin flip and conserve total spin angular momentum, and $f(\epsilon)$ is the Fermi-Dirac distribution function.  The change in energy of the nuclear spin is essentially zero so  $\epsilon_i = \epsilon_f$, but the excitations can come from  nodes at different positions, ${\bf k}_{i,f}$, on the Fermi surface.  Arguments extending this framework to the superconducting state involve coherence factors\cite{mac76} important in elemental BCS superconductors, but which have been irrelevant to the class of HTS superconductors, largely owing to strong anisotropy.  Furthermore, for exotic or unconventional pairing systems there is an interplay between order parameter structure, impurity scattering, and high magnetic fields, together with the shift of the local electronic energy spectrum from the Doppler effect, which requires sophisticated analysis.  One limit that is arguably tractable is that of low temperatures for nodal superconductors where the excitations are Dirac-like.  In this case the thermal contributions to the integral in Eq.~1 are less important and limiting forms of $N(\epsilon)$ at low energy  can be used.  For cuprates this is simply,  $N(\epsilon) \propto \epsilon$.  

It is a common error in interpretation of \rt\, in the superconducting state to look for power law temperature dependences below $T_c$, say in the region $ \sim$$0.5 <T/T_c< 1.0$ where there is a sudden drop in the relaxation rate compared to the normal state of the metal.  This type of analysis corresponds to an inappropriate expansion of Eq.~1 in a region in which the structure of the order parameter has not approached its low temperature limiting form.   Impurity effects on $N(\epsilon)$ are difficult to quantify and, in principle, cannot be excluded even for nominally clean single crystals of superconductors.  However, the low temperature limit is potentially accessible in clean samples at sufficiently high magnetic fields such that the field energy is larger than the impurity bandwidth. 

Experimental evidence for spatially dependent relaxation came from \Ox\, NMR in isotopically enriched, optimally doped, aligned powders of \YBCO.\cite{cur00,mit01b}  In the first experiments by Curro \et,\cite{cur00} the central transition of the planar oxygen was isolated from the overlapping apical oxygen resonance by a technique of population-enhanced, double resonance.\cite{haa98} The relaxation rate was measured at different points in the NMR spectrum in a magnetic field of $H = 8.3$ T at several temperatures below the vortex melting transition temperature. The relaxation was found to be inhomogeneously distributed and increasing with $H_{int}$, {\it i.e.} with frequency within the NMR  spectrum for the planar \Ox\, site.  At temperatures below $T_c$ the apical \rt\, has a $T$-linear temperature dependence, whereas the planar \rt\, is $\propto T^3$ upon entering the superconducting phase, but becomes \rt\, $\propto T$ at low temperatures. These authors also measured \rttwo\, and both rates were interpreted as evidence of vortex vibrations for $T < 25$ K and from quasiparticle scattering for $T \geq 25$ in the vortex solid state. However, theoretical investigations  of the effect of vortex vibrations on \rt\, by Wortis \et\cite{wor00} have suggested that this interpretation is unlikely.
\begin{figure*}[!]
\centering
	\includegraphics{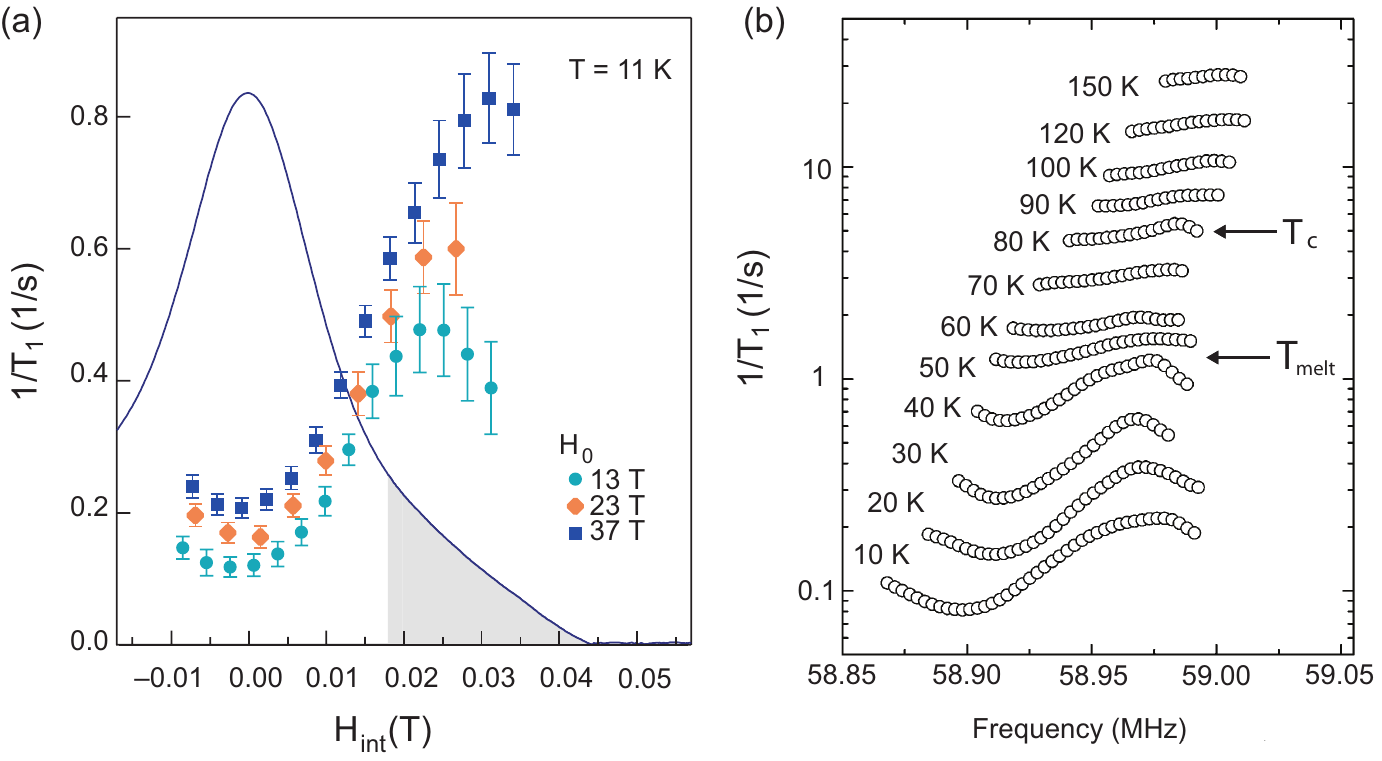}
	\caption{(a) The frequency dependent \Ox\, relaxation at various magnetic fields in \YBCO\, from Mitrovi{\'c} \et\cite{mit01b} ; and (b) for YBa$_2$Cu$_4$O$_{8}$ at various temperatures from Kakuyanagi \et\cite{kak02}  For  \YBCO,  panel (a), as the external magnetic field is increased the relaxation rate at each frequency position in the spectrum shows a constant increase attributed to the Zeeman effect.  Additionally, \rt\, increases with  frequency across the spectrum consistent with a Doppler effect.  The shaded region corresponds to the fraction of the measured \Ox\, spectrum, shown in the background, that would correspond to nuclei in the vortex core for $H=37$ T.   For panel (b), when YBa$_2$Cu$_4$O$_{8}$ is cooled through the superconducting transition, the relaxation shows no frequency dependence as would be expected for a motionally averaged vortex liquid.  However, as the sample is cooled below the vortex freezing temperature, the relaxation becomes frequency dependent, characteristic of Doppler shifted  relaxation with the exception of the highest and lowest frequency portions of the spectrum. The two experiments, (a) and (b),  have clear similarities.}
	\label{SDW1}
\end{figure*}

Mitrovi{\'c} \et\cite{mit01b,mit03} reported results on isotopically enriched \Ox\, optimally-doped, aligned powders of \YBCO. Their work was performed up to very high magnetic fields,  $H \leq 42$ T, Fig. \ref{SDW1}(a).  These authors came to a different conclusion than Curro \et\cite{cur00} regarding the low temperature mechanism for relaxation.  Mitrovi{\'c} \et\,  examined the -1/2~$\leftrightarrow$~-3/2 transition to separate the planar and apical oxygen resonances. The temperature and magnetic field dependence of \rt\, relaxation was analyzed in terms of scattering of quasiparticles between the gap nodes on the Fermi surface.  The key features of their experiments are the dependence of \rt\, on magnetic field and extension of measurements to very high fields to permit identification of the relaxation mechanism and sensitivity to vortex core excitations.  At low temperatures the NMR spin-lattice relaxation rate due to quasiparticle scattering is given by,\cite{mit01b}
\begin{equation}
T_1^{-1} \approx \langle|\epsilon - Z +D_i||\epsilon+Z+D_f|\rangle
\end{equation}
where $\epsilon \approx k_BT$ is the thermal contribution, $Z = \gamma_e\hbar H/2$ is the Zeeman energy, and $D_{i,f} =  ({\bf v}_F)_{i,f}$$\cdot$${\bf p}_s$ are the initial and final Doppler shifts.  By changing the external magnetic field, temperature, and position in the NMR spectrum, each of these terms changes magnitude.

\begin{figure}[t]
\centering
	\includegraphics[width=.45\textwidth]{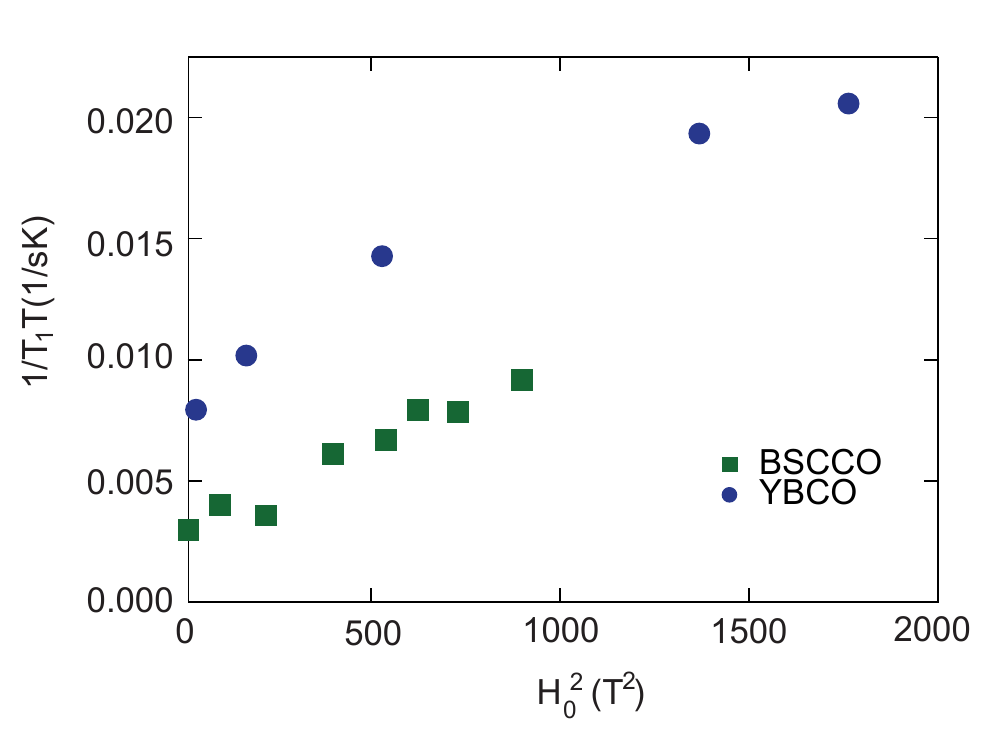}
	\caption{The magnetic field dependence of $^{17}$O \rt\, for two $d$-wave superconductors selectively measured at the saddlepoint in the magnetic field distribution where the Doppler terms mostly cancel by symmetry.  Here there is evidence for a Zeeman contribution to the quasiparticle excitations which vary with magnetic field according to $\propto$ $H^{2}$, reproduced from Oh \et\cite{oh11b}  The zero field limit of \rt\, corresponds to thermal relaxation and non-magnetic impurity effects, surprisingly similar in these two different superconductors.  The slopes for the two materials would be the same if the electronic $g$-factors were the same.} 
	\label{zeeman}
\end{figure}

 With increasing magnetic field, \rt\, increases quadratically with $H$ at the saddle point of the spectrum where D is minimal.    In this scenario  \rt $ \sim |Z^2-\epsilon^2|$, and at low enough temperature becomes  $\sim |Z^2|$, since the thermal term varies as $T^3$.  This behavior is demonstrated in Fig. \ref{zeeman} that compares the behaviors of \YBCO\, and \BSCCO\, showing an increase in \rt\, $\propto$ $H^{2}$.  At locations where the Doppler term is more prevalent and at low temperatures, \rt\ is given by $|\pm D^2 -Z^2|$ and the sign of D depends on the nodes for initial and final quasiparticle states.  Observation of an increasing \rt\, with magnetic field suggests\cite{mit01b} that internode scattering dominates, ${\bf k}_{i} \neq {\bf k}_{f}$.  
 
Experiments by Kakuyanagi \textit{et al.},\cite{kak02} Fig. \ref{SDW1} (b), report similar effects.  The authors concentrate on the non-monotonic variations of \rt\, across the spectrum. Separating the planar and apical resonances by examining the  -1/2 $\leftrightarrow$ -3/2 transition, they found different behavior at the different oxygen lattice sites.  The planar oxygen shows an increase in \rt\, across the spectrum which reaches a maximum and then diminishes at high frequency, approaching the vortex core.  The apical oxygen shows a  more slowly increasing \rt, with a magnitude 5 times less than the planar site.  The different behavior and magnitude of \rt\, at these sites seems to indicate a different mechanism between the two locations, specifically these authors conclude that the planar oxygen \rt\, is dominated by quasiparticle interactions.  They argue that if there were to be vortex vibrations they would affect the two sites in the same way.  The dip in \rt\, at high frequency, a characteristic inconsistent with Doppler shift, was interpreted as an effect of the vortex core, where superconductivity is suppressed, Fig. \ref{SDW1} (b). Also, an increase on the low frequency side of the spectrum could not be explained in terms of Doppler effects.  There is a striking similarity in the two measurements of \rt\, on \YBCO\, and YBa$_2$Cu$_4$O$_{8}$  for comparable magnetic fields, $H \lesssim 10$ T.

This body of experimental work paints a reasonably consistent picture. Certainly \rt\, increases across the NMR spectrum due to vortices, and another contribution other than the Doppler shift affects \rt\, near the vortex core, corresponding to the high frequency tail of the spectrum. The small upturn in \rt\, on the low frequency part of the spectra is  a puzzle, and appears to be ubiquitous to all reports to date: \YBCO,\cite{mit01b,mit03b} YBa$_2$Cu$_4$O$_{8}$,\cite{kak02} Tl$_2$Sr$_2$BaCu$_2$O$_{6.8}$,\cite{zhe02} \TBCO,\cite{kak03}  and for \BSCCO.\cite{mou11b}  However, theoretical input helps clarify possible mechanisms for the inhomogeneous relaxation and further detailed experiments in the vortex core region of the spectrum will be essential to reveal the spin character of the vortex cores in HTS.

\begin{figure}[!]
\centering
	\includegraphics{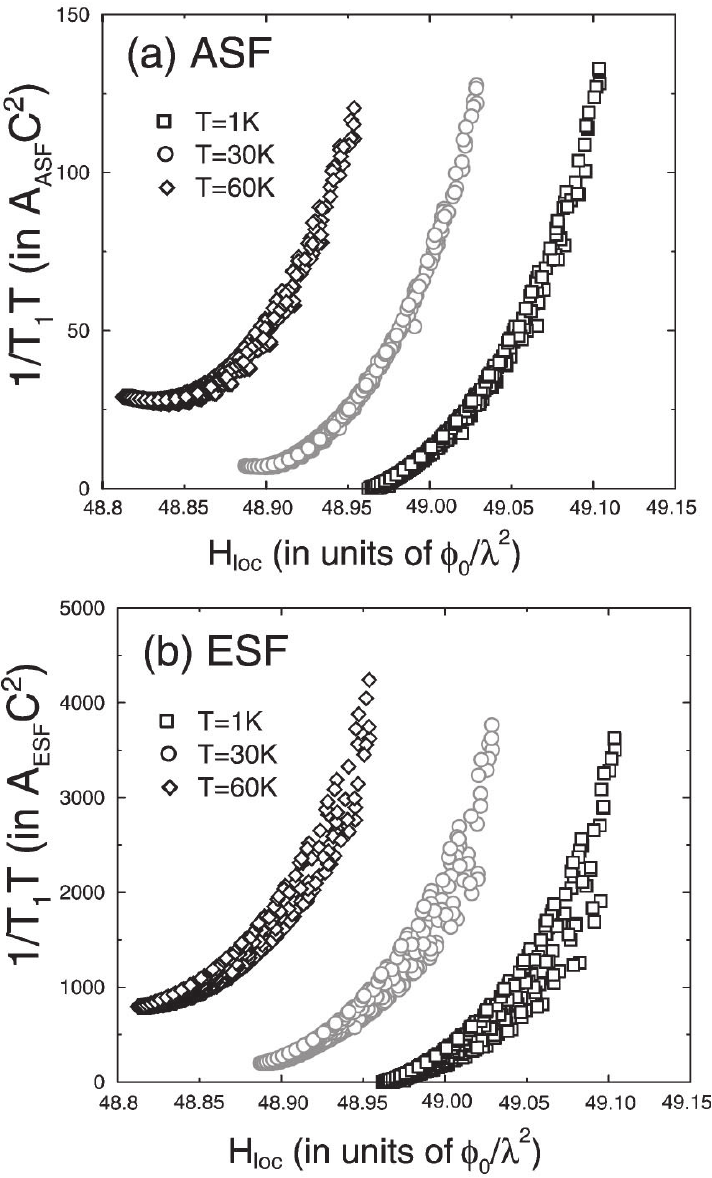}
	\caption{The local magnetic field dependence of \rt\, at different temperatures $T < T_m$ from Morr's theory.\cite{mor01}  Bottom panel, intra-node scattering (electron spin flip, ESF).  Top panel, inter-node scattering (antiferromagnetic spin-flip, ASF) processes.  In the top panel, ASF scattering shows a non-monotonic increase of \rt\, across the NMR spectrum causing a local minimum at the saddle point.  In contrast the ESF scattering has a monotonic temperature dependence of \rt\, such that it increases with temperature and frequency at all points in the spectrum.  The curves are offset for clarity.}
	\label{ESFvsASF}
\end{figure}
 
 \begin{figure}[t]
\centering
	\includegraphics{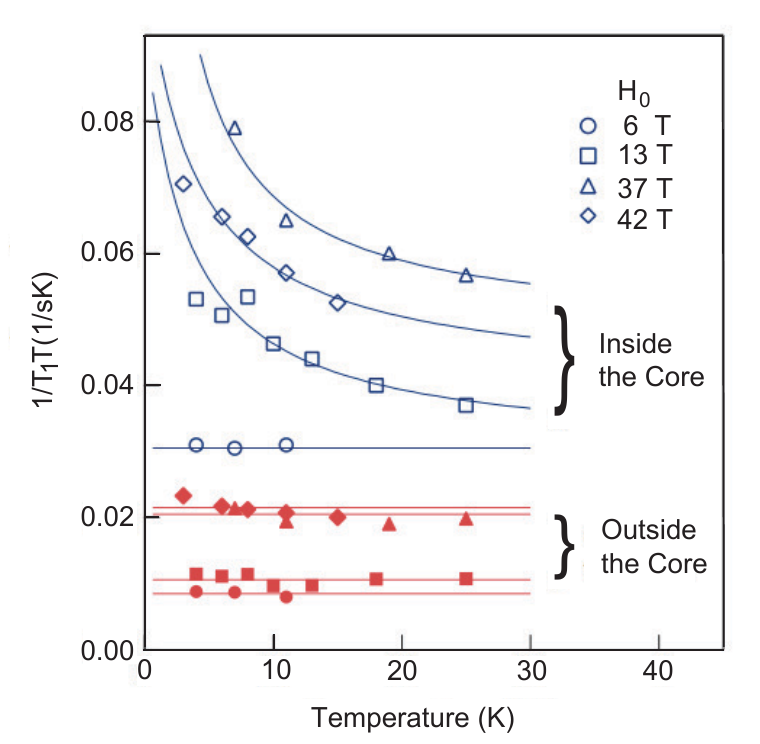}
	\caption{\rt\,  temperature dependence at different magnetic fields and vortex lattice locations.   \rtt\, was found by Mitrovi{\'c} \et\cite{mit03} to be a constant outside the vortex core in \YBCO.  At $H = 6$ T, \rtt\, is constant presumably because of insensitivity to the core at low magnetic field; however, with increasing $H$ the temperature dependence follows a $1/(T-\theta)$, behavior of a Curie-Weiss law indicative of antiferromagnetism.  }
	\label{T1core}
\end{figure}

\subsection{Spin-Lattice Relaxation: Theory}

Early experimental indications of spatially inhomogeneous relaxation in \YBCO\, quickly spurred several theoretical models.  The two \rt\, mechanisms are the effect of vortex vibrations and the spin flip scattering of quasiparticles. Wortis \et\,\cite{wor00}  compared the relaxation due to these two processes for $^{63}$Cu. Their calculations resulted in spatially inhomogeneous relaxation which increased with internal magnetic field for both mechanisms, however the calculated \rt\, due to vortex vibrations is much too slow to account for experimental results.  


There are two regimes examined for spin flip scattering of
quasiparticles; the weakly interacting electronic spin flip scattering
(ESF) and the strongly interacting antiferromagnetic spin fluctuations
(ASF) both of which show a frequency dependent \rt.  The effects on both
copper\cite{wor00} and oxygen\cite{mor01,kna02} of
ESF, where quasiparticles scatter onto the same node on the Fermi
surface effectively flipping spin but maintaining momentum,  were found to have a
uniform change of \rt\, with temperature, Fig. \ref{ESFvsASF}(b).
Additionally, the Doppler shift is related to the angle between the
scattering wave vector and the underlying crystal lattice giving a
much larger distribution in \rt.   In the strongly interacting ASF
limit, Fig. \ref{ESFvsASF}(a), the quasiparticles  scatter
between different nodes of the Fermi surface with a large change in
momentum, and a sign change in the Doppler term of Eq.~1. This can
give a non-zero value of relaxation at the minimum frequency point of
the spectrum and a nonuniform change in \rt\, with temperature.  It
was later determined\cite{kna02} that, in the quantum limit, ESF can
also cause a non-monotonic \rt\, such that there is a local maximum in
the rate at the minimum field point with the requirement that the line
from the vortex along the node direction must cross the minimum of
$H_{int}$.

A different approach based on finding analytical and numerical solutions of the Bogoliubov-de Gennes equations was taken by Throckmorton and Vafek\cite{thr10} calculating the shifts and relaxation including strong magnetic fields up to 42 T.  Their results show that in the near vortex core region antiferromagnetic fluctuations are not necessary to account for experiment.\cite{mit01b,mit03} Rather they suggest that quasiparticle pair creation and annihilation (PCA) can be a dominant  contribution to \rt, especially important for high field NMR.  
 
In general the theory captures the experimental trends.  However, there remain several questions at this point. Is the behavior of \rt\, at the lowest frequencies in the spectrum a manifestation of the ASF model?  Can the theory be extended to high magnetic fields and incorporate the Zeeman interaction for quasiparticles?  And finally what is expected near the vortex core and is the PCA mechanism the dominant one?

\subsection{Vortex Cores}

With the spatial resolution of \rt\, established, the vortex core can be probed if the magnetic field is sufficiently large for NMR signal to be extracted from the nuclei that occupy the area within the vortex cores. The inter-vortex spacing, $d$, scales with magnetic field as $d = \sqrt{\phi_0/H}$.  Taking the vortex core radius to be $\xi$, the vortex core will occupy a fraction, $\sim\pi\xi^2/d^2$, of the cross-sectional area of the sample, which is the same fraction of the NMR spectrum that can be associated with nuclei in the core, the highest frequency part of the spectrum shown shaded in Fig.~\ref{SDW1}. This leads to a field dependent vortex fraction $A_{vortex}/A_{total} = \pi H\xi^2/\phi_0$, a quantity that grows linearly with external magnetic field.

The spin 1/2 nucleus $^{205}$Tl has been used by Kakuyanagai \et\cite{kak03} to probe the vortices in \TBCO.  The $^{205}$Tl nucleus, although not in the superconducting plane, is sensitive to antiferromagnetic fluctuations from the $^{63}$Cu sites.  By comparison, \Ox\, is less sensitive to antiferromagnetic fluctuations owing to cancellation of the singlet correlations between neighboring Cu.  The $^{205}$Tl NMR measurements of \rt, performed in low magnetic field, $H = 2.1$ T,  increased  quite steeply with increasing frequency from the saddle point field. Spectra at different temperatures were discussed in terms of the full width at half maximum, $\delta$f, and the frequency width between points  at 1\% of the peak in the spectrum, $\Delta$f.  Below T = 20 K, $\Delta$f becomes much broader than expected from a London model while $\delta$f changes rather little.  Furthermore, \rt\, is enhanced in the high frequency region with a peak at $T = 20$ K,  taken to be the Ne\'el temperature of the vortex core.

\begin{figure}[b]
\centering
	\includegraphics[width=.45\textwidth]{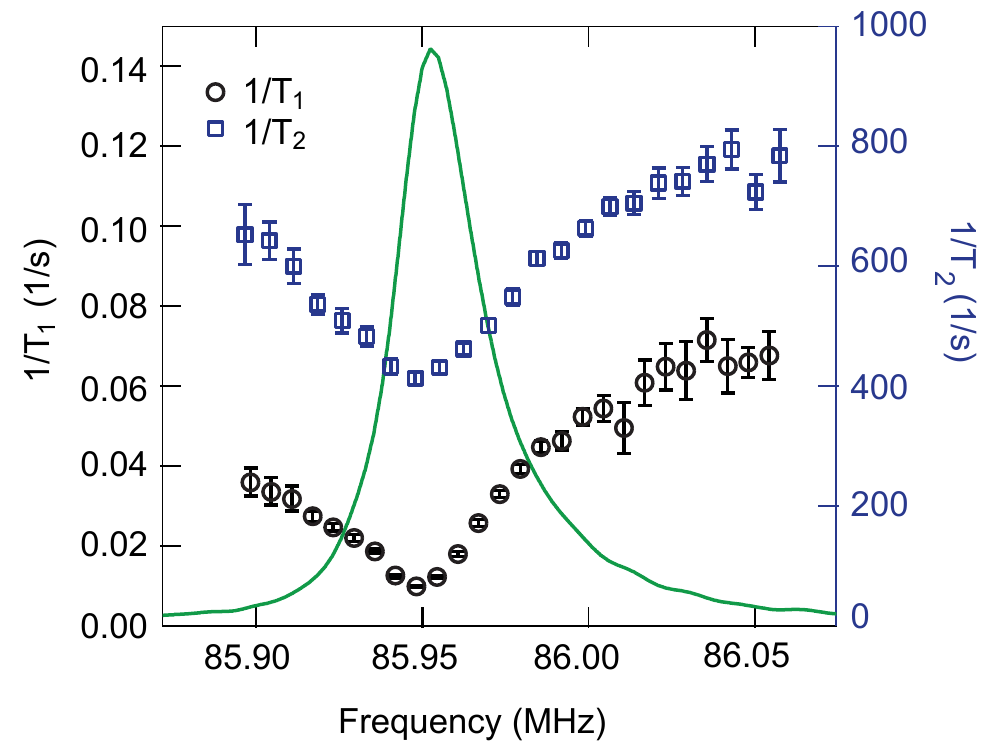}
	\caption{The frequency dependent \Ox\, relaxation for \BSCCO\, at $T = 4$ K and $H = 15$ T from Mounce \et\cite{mou11b}  The spin-lattice relaxation rate (black,circles), \rt\, and the spin-spin relaxation rate (blue, squares), \rttwo, have a strong non-monotonic relaxation as a function of frequency contrary to what is seen in \YBCO\, and \TBCO.  This behavior can be explained by a  spin-density wave induced by the vortex core in its near neighborhood over a distance $\sim 2\xi$.  In this model, the frequency distribution in the spectrum is relabeled non-monotonically, oscillating with position approaching the vortex core.}
	\label{SDW2}
\end{figure}

Mitrovi{\'c} \et\cite{mit03} have exploited availability of very high magnetic fields at the NHMFL to reach 42 T.  These measurements have a  substantial fraction of the \Ox\, NMR spectrum of \YBCO\, in the vortex core, nearly 16\% of the total spectrum.  The temperature dependence of \rtt\, at the saddle point were compared with that in the vortex core as shown at several magnetic fields in Fig. \ref{T1core}.  Outside the core \rtt\, is a constant, but in the vortex core region, in magnetic fields above $H = 6$ T, \rtt $\propto 1/(T-\theta)$ following a Curie~-~Weiss law similar to $^{63}$Cu relaxation in the normal state, however with $\theta \lesssim 0$.  This was interpreted as evidence of strong antiferromagnetic fluctuations in the vortex core. 

\begin{figure*}
\centering
	\includegraphics{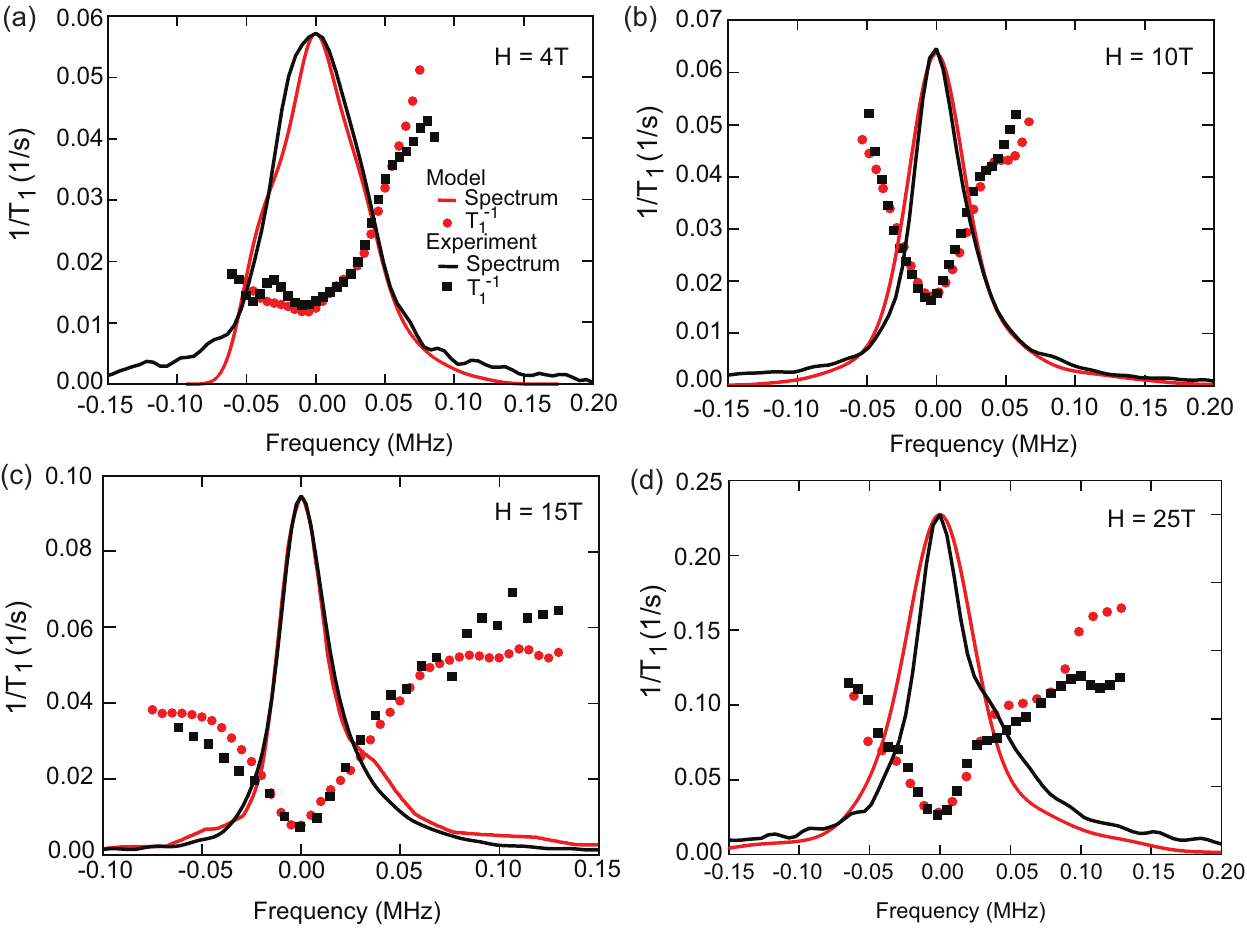}
	\caption{The  spectrum and  frequency dependent \rt\, relaxation rate at T = 4 K in \BSCCO\, from Mounce \et\cite{mou11b} for experiment (black) and their model calculation (red).  With increasing external magnetic field, the amplitude of the SDW component of the spectrum increases relative to the contribution from supercurrents.  This leads to an increase of \rt\, at the lower frequency portion of the spectrum. }
	\label{SDW5}
\end{figure*}

\subsection{Spin-spin relaxation: spatial resolution}

Although the spin-spin relaxation, \rttwo, has not gained as much attention as \rt, it has also been  shown to have a spatial dependence.\cite{cur00,mou11b} Specifically,  \rttwo\, in both \YBCO\, and \BSCCO\, follow similar  frequency dependence at low temperature as \rt.   A direct comparison of \rt\, and \rttwo\, in Fig.~\ref{SDW2} emphasizes their similarity and naturally suggests that there are at least two significant contributions to \rttwo, one of which has the same origin as for \rt\, and which we believe depends on vortex supercurrents. The second is independent of frequency.  The unusually pronounced upturn in the rates at low frequency in the NMR spectrum will be discussed in the next section.  It was also shown that \rttwo\, in \YBCO\, has similar behavior at different crystallographic positions, $^{89}$Y and both \Ox\, sites, suggesting that the relaxation has a common origin likely determined by vortex vibrations.\cite{rec97,cur00} Although the theory\cite{lu06} for vortex dynamics is not settled, as we shall discuss next, nonetheless its affect on \rttwo\, is predicted to depend strongly on position in the vortex unit cell quite similar to calculations for \rt.   

Calculations\cite{lu06} based on Langevin dynamics of a frequency dependence of \rttwo\, due to over damped vortex vibrations show qualitatively the same behavior as \rt. However, the calculated rate is 3 orders of magnitude slower than what is experimentally observed.  Furthermore, the spin-echo decay rate follows neither a gaussian decay as expected for a $^{63}$Cu \rt\, mechanism\cite{wal95} or an exponential decay due to vortex dynamics in the solid state.\cite{bac98}  Although there is evidence that vortices play a role and that \rttwo\, is spatially resolved, further work will be required to establish the basic mechanisms that are responsible for \rttwo.

\subsection{SDW in \BSCCO}

The spatial dependence of \Ox\, relaxation  in \BSCCO\, single crystals has been found to be significantly different than in \YBCO.\cite{mou11b} Both \rt\, and \rttwo\, show a non-monotonic correspondence between relaxation and local magnetic field, Fig.~\ref{SDW2}, much more so than can be accounted for by any combination of Doppler shifts and broadened Redfield patterns for the distribution of local fields.  It has been suggested that antiferromagnetic polarization near the vortex core could disturb the distribution of local magnetic fields and broaden the NMR spectrum.\cite{mit03b,thr10} Calculations along these lines were performed by these authors  to account for broadening observed in the \YBCO\, aligned powder experiments.  According to one model,\cite{mit03b} the magnetic moment in the vortex core required to account for the broadening would need to be 2-5 $\mu_B$ if the antiferromagnetic wave vector is taken to be $\sim \pi/a_0$, where $a_0$ is the crystal lattice constant.  However, the size of this moment is unreasonably large.\cite{mitthesis}

There is independent evidence for vortices developing spin-polarization around the vortex core. Indirectly, STM experiments in \BSCCO\, have shown a "checkerboard" pattern of the local density of states with periodicity 4$a_0$ which should be associated with an 8$a_0$ spin density modulation.\cite{hof02,han04, wis08}  And there is direct evidence from neutron, elastic scattering experiments, performed on (La$_{1-x}$Sr$_x$)$_2$CuO$_4$ which  indicate magnetic Bragg scattering  with a wavevector of $\pi/8a_0$.\cite{lak01, lak02, kha02}  Furthermore, there have been theoretical predictions with models involving magnetic competition and/or coexistence with superconductivity consistent with these results.\cite{sac03}

The magnetic field dependence of the spectral linewidth  at $T = 4$ K in clean single crystals of  \BSCCO\, is well beyond what might be expected from a GL-calculation of the NMR spectrum. A phenomenological model of the local magnetic fields was suggested by Mounce \et,\cite{mou11b} including contributions from vortex supercurrents superposed on a damped spin density modulation centered at the vortex cores,

\begin{equation}
B (x,y) = A\mathrm{cos}(2\pi x/\lambda)\mathrm{cos}(2\pi y /\lambda)e^{-(x^2+y^2)/\sigma^2}
\end{equation}

\noindent
Here A is the amplitude of the paramagnetic contribution, $\lambda$ is the periodicity of the modulation and $\sigma$ is the characteristic decay length away from the vortex core. Fitting these to the distribution of local magnetic fields results in $A$  increasing approximately linearly with magnetic field while $\sigma$ is roughly constant and given by $\sim 2\xi$.  However, the addition of an oscillating local magnetic field, Eq.~3, dramatically changes the dependence of \rt on frequency throughout the NMR spectrum, Fig.~\ref{SDW2} and \ref{SDW5}.  At low applied fields the relaxation profile is similar to that found in \YBCO\, as there is a correspondingly small contribution from vortex core, spin polarization.  As external magnetic field increases, the spin density wave amplitude increases while the vortex supercurrent component decreases producing the non-monotonic behavior.  In this work the relaxation rate was attributed to Doppler shifts of the nodal quasiparticles and was taken to be proportional to the square of the supercurrent momentum in the CuO$_2$ plane, calculated as in Fig.~\ref{specandps}.

\subsection{$s$-wave superconductors}

Without nodes in an energy  gap at the Fermi surface, the quasiparticle bound states in an $s$-wave superconductor are localized in the vortex core. The energy of the core states were calculated to be $E_\mu = \mu\Delta_0^2/ E_F$, with $\mu$=1/2, 3/2... , relative to the Fermi energy and are important at low temperatures where they are evident in the zero-bias anomaly in tunneling experiments.\cite{car64,sho89} The spatial dependence of \rt\,  would be homogeneous with the exception of the vortex core unless spin diffusion were to play a particularly important role; see Appendix B. 

\begin{figure}[t]
\centering
	\includegraphics[width=0.4\textwidth]{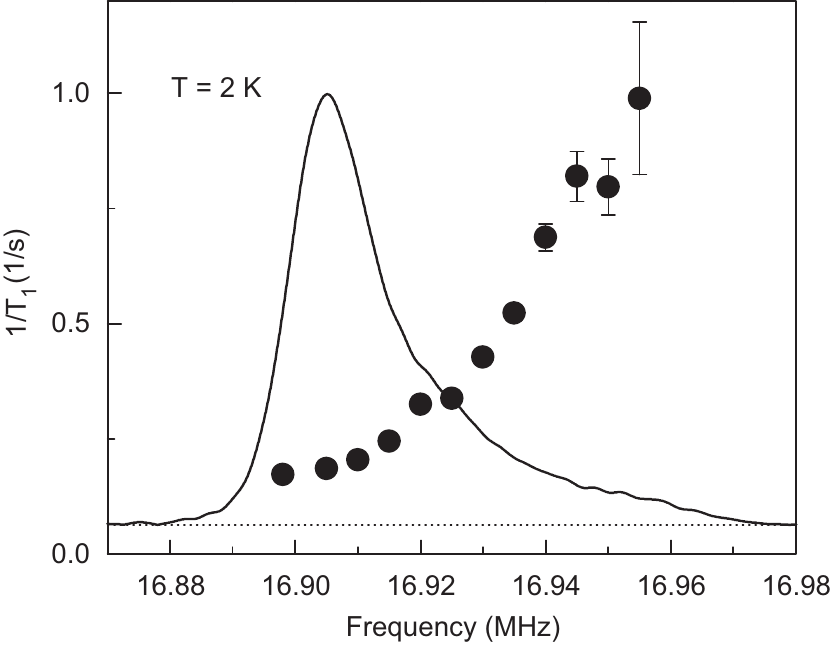}
	\caption{The NMR spectrum and relaxation, \rt\, for the skutterudite compound, LaRu$_4$P$_{12}$  in relatively high magnetic field $H= 0.98$ T, $H/H_{c2} = 0.31$, reported by Nakai \et\cite{nak08}}
	\label{skud}
\end{figure}

\begin{figure}[b]
\centering
	\includegraphics[width=.45\textwidth]{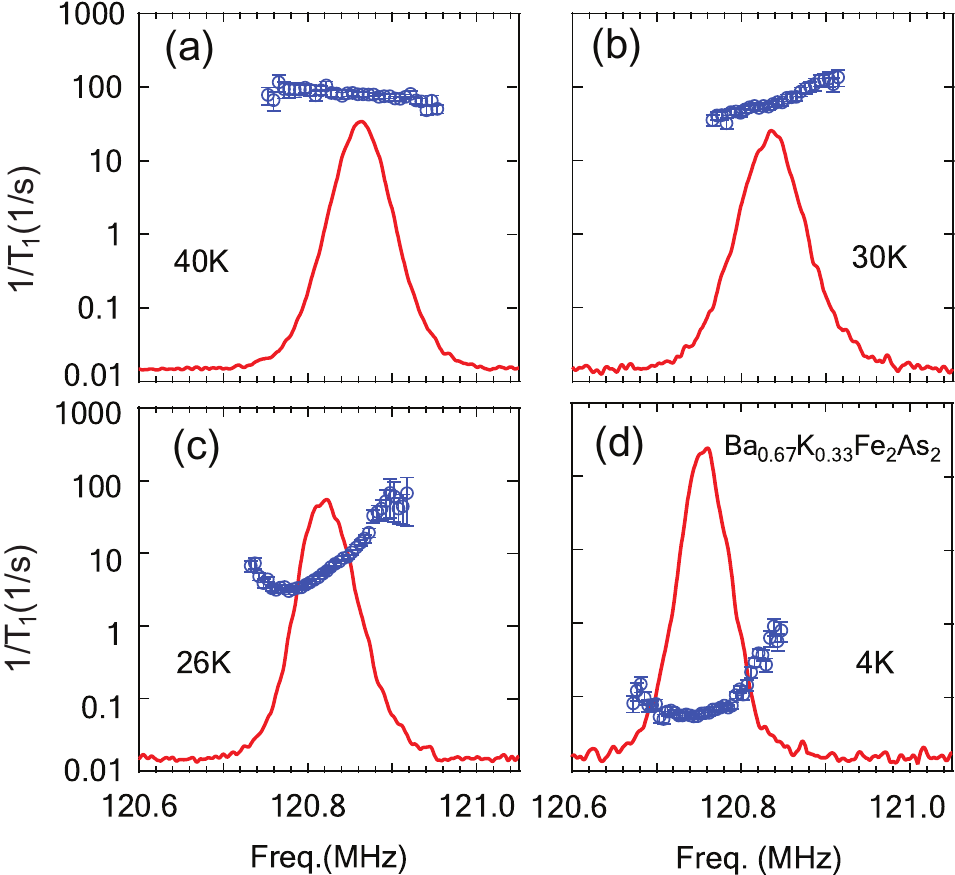}
	\caption{The frequency dependent $^{75}$As, relaxation for Ba$_{0.67}$K$_{0.33}$Fe$_2$As$_2$, at  $H = 16.5$ T from Oh \et\cite{oh11b}  At $T=4$ K, the spin-lattice relaxation rate  has a strong frequency dependence across the NMR spectrum varying by one order of magnitude increasing with increasing frequency.  This profile develops only in the superconducting state where $T_c= 38$ K.}
	\label{BaK122}
\end{figure}

Recent NMR in the $s$-wave superconducting skutterudite LaRu$_4$P$_{12}$ ($T_c = 7.2$ K) has a frequency dependent relaxation.\cite{nak08}  Above $T_c$ the $^{31}$P NMR spectrum is symmetric with a constant \rt\, throughout the spectrum, Fig. \ref{skud}.  As it is cooled below $T_c$ the spectrum becomes asymmetric with the familiar Redfield-like pattern, and \rt\, develops a profile with a monotonic increase at higher frequencies.  Since there are no nodal quasiparticles, the distribution of \rt\, was attributed to diffusion from the vortex core.  This is reasonable according to the theory of spin diffusion since  these effects are larger for high gamma nuclei varying as the local dipolar field squared, $H_d^2 \sim \gamma^2$.  For  $^{31}$P this is about an order of magnitude larger than it would be for \Ox.  The experiments are done in relatively high magnetic field, $H/H_{c2} = 0.31$ where the vortex cores strongly overlap and half of the probe nuclei are within two coherence lengths of the core.

The recently discovered pnictide high $T_c$ superconductors\cite{kam08} are thought to have $s\pm$ superconducting gap symmetry, a generically multiband superconductor with two principal gaps, one with hole excitations and the other around an electron pocket.\cite{maz08} A calculation by Bang\cite{ban10} indicates  that the density of states will produce a Volovik-like effect where the quasiparticles have a finite density of states outside the vortex cores in a region where the supercurrents are sufficiently high to destabilize Cooper pairing from the smaller gap.  It is especially important to include the interplay between interband scattering from impurities, temperature and magnetic fields.   Bang's theory finds that the density of states has a linear field dependence at low magnetic fields, which a naive interpretation would suggest must produce a quadratic field dependence for the spatially averaged, \rt\,$\propto H^2$. However, spin-lattice relaxation in the mixed state in a multiband superconductor is not so straightforward and a more careful analysis\cite{ban11} leads to a linear field dependence.  Recent measurements on Ba$_{0.67}$K$_{0.33}$Fe$_2$As$_2$ of  \rt\, as a function of local magnetic field have revealed\cite{oh11b} an inhomogeneous distribution of relaxation that appears only in the superconducting state and this behavior is quadratic in field.  In Fig.~\ref{BaK122} we show the evolution of the \rt\, profiles as a function of temperature providing evidence that Doppler effects play a role.  However, discrepancy with the theory\cite{ban11} is not understood. 

\section{Conclusion}

This brief survey focuses on NMR in high temperature superconductors emphasizing the effects of vortex supercurrents on the spectrum and their Doppler effects on quasiparticle excitations and correspondingly on spin-lattice and spin-spin relaxation. At this stage of development in the field it is clear that frequency resolved relaxation spectroscopy has important contributions to make to understanding vortex structures in unconventional superconductors.  Frequency dependence implies spatial organization since the vortex currents increase in amplitude approaching the vortex core $\propto 1/r$, a consequence of the London equations.  At the same time it is also clear that interpretation of experiments is greatly facilitated if samples are high quality single crystals with well-established electronic properties.  Measurements at very high magnetic fields have been valuable and will continue to be significant particularly to improve our knowledge of vortex core excitations with a spin-dependent probe.  In this regard NMR has a unique role to play.  Our review discusses the subject of NMR investigations of vortices, mainly from the authors' perspectives and their contributions to it.\\

\section{Acknowledgements}

We thank Nate Bachman, Yunkyu Bang, Pengcheng Dai, Matthias Eschrig, Morten Eskildsen, Kasu Fujita, Yuji Furukawa, Philip Kuhns, Moohee Lee, Vesna Mitrovic, Bill Moulton, Sutirtha Mukhopadhyay, Arneil Reyes,    Eric Sigmund, and  Anton Vorontsov for their contributions. Support for our work is acknowledged from the U.S. Department of Energy, Office of Basic Energy Sciences, Division of Materials Sciences and Engineering, award DE-FG02-05ER46248.

\section{Appendices}

\subsection{Volovik Effect}
For a simple metal \rt\, is expressed in terms of the thermal average of the joint density of states:
\begin{equation}
1/T_1 \propto  \int N(\epsilon_i)N(\epsilon_f) f(\epsilon_i)[1-f(\epsilon_f)]\, \mathrm{d} \epsilon
\end{equation}
where $i$ and $f$ indices label the initial and final electron states,  $f(\epsilon)$ is the Fermi-Dirac distribution function, and  $\epsilon_i \approx \epsilon_f$.\\

In a spin-lattice relaxation experiment we measure the NMR nuclear magnetization $M(t)$ as a function of the time during relaxation after an $RF$-excitation that disturbs $M$ from equilibrium. Since the mixed state of a superconductor is spatially inhomogeneous we are concerned with the spatially averaged NMR response.  The resulting $M(t)$ is the superposition of the responses of nuclei from all regions of the sample. The initial time evolution of the magnetization, $i.e.$ the limit as $t\rightarrow 0$ of $M(t)$, is equivalent to the magnetization evolution at the average rate.  This allows us to compare experiment with a theoretical calculation  of the spatially averaged rate obtained from Eq.~4, $i.e. <1/T_1>_{av}$.
\begin{equation}
<1/T_1>_{av} \,\, \,\,\propto\,\, \, \int <N(\epsilon_\uparrow)N(\epsilon_\downarrow)>_{av}\, f(\epsilon_\uparrow)[1-f(\epsilon_\downarrow)]\, \mathrm{d} \epsilon
\end{equation}
where the integral is a thermal average over all possible quasiparticle states.
\noindent
Therefore we focus on just the important part of the integrand,
\begin{equation}
<N(\epsilon_\uparrow)N(\epsilon_\downarrow)>_{av}\,\, \,\, \propto\,\, \,\,\frac{1}{d^2} \int_\xi^d \! (N[\epsilon(r)])^{2}\,r\,\mathrm{d}r
\end{equation}

\noindent
where $\epsilon(r)$ is the spatial dependence of the energy in the vortex unit cell of size $d$ and the integral goes from  the vortex core radius, $\xi$, to $d$.  If the temperature is sufficiently low then the relevant quasiparticle excitations are small,  we can expand $N(\epsilon)$ near $\epsilon=0$:  $N(\epsilon) \sim \epsilon^n$, provided that the density of states itself is not modified approaching a vortex core.  For example, in the case of a clean  $d$-wave superconductor the nodal quasiparticles have a dispersion, $n=1$.

Now we introduce the inhomogeneity of the mixed state owing to the Doppler shift of the elementary excitations, $ \delta \epsilon \sim {\bf v}_F \cdot {\bf p}_s$, where ${\bf p}_s\, \propto \,1/r$.
\begin{equation}
<N(\epsilon_\uparrow)N(\epsilon_\downarrow)>_{av}\,\,\,\,  \propto\,\,\,\, \frac{1}{d^2} \int_\xi^d \! r^{-2n}\,r\,\mathrm{d}r
\end{equation}

\noindent
For a $d$-wave superconductor $n=1$, $i.e.$  $N(\epsilon)\,\, \,\,\propto\,\,\,\, \epsilon$, and the spatially averaged rate is given in Eq.~8 where we note that $Hd^2=\phi_0$, and $\phi_0$ is the flux quantum:
\begin{equation}
<1/T_1>_{av-d}\,\, \,\, \propto\,\, \,\,\frac{1}{d^2} \,\ell n \frac{d}{\xi}\,\, \,\, \propto\,\, \,\,H\,\,\ell n \frac{H_{c2}}{H}\,\,\, \approx \,\,\,H.
\end{equation}

\noindent
This is the accepted behavior for the spin-lattice relaxation rate for a $d$-wave superconductor and was discussed by Volovik\cite{vol93} for the spatially  averaged density of states in the mixed state.

Also for the $d$-wave superconductor we can write the average density of states as:
\begin{equation}
<N(\epsilon)>_{av-d}\,\,\,\,  \propto\,\,\,\, \frac{1}{d^2} \int_\xi^d \! r^{-1}\,r\,\mathrm{d}r\,\,=\,\,\frac{1}{d^2}(d - \xi)
\end{equation}
\begin{equation}
<N(\epsilon)>_{av-d}\,\,\,\, \approx\,\,\, \frac{1}{d} \,\,\,  \propto \,\,\,H^{1/2}
\end{equation}

\noindent
as expected.  However, the approximations above and the conclusions reached are not valid for a $s\pm$-superconductor.\cite{ban11}

On a historical note, a calculation along these lines was first made by Volovik\cite{vol88} in 1988 in the context of encouraging experimental efforts to explore the symmetry of the order parameter of the heavy fermion compound UPt$_3$.  The Doppler effect from vortex supercurrents was then discussed theoretically by Yip and Sauls\cite{yip92} in 1992 for the $d$-wave case and then again by Volovik\cite{vol93} for the $d$-wave superconductor in his now famous paper in 1993.

Finally, to emphasize a technical point, the average of \rt\, is not equal to a rate determined from a standard relaxation experiment where the entire NMR spectrum is excited and detected.  In fact, only the initial part of the recovery of the relaxation profile will correspond to the average rate, as we noted above.  Alternatively, one can measure the frequency resolved rates across the spectrum and then calculate $<T_1^{-1}>_{av}$ as a spectrum-weighted average.

\subsection{Spin Diffusion}

NMR measurements on superconductors and their interpretation include early work on the mixed state of low temperature superconductors, notably vanadium compounds, where \rt\,  was thought to be inhomogeneously distributed owing to spin diffusion.\cite{sil66,sil67}  The argument is that localized quasiparticle excitations in the vortex core, $i.e$ bound states,\cite{car64} would be a source of relaxation which, through nuclear spin diffusion from outside the core region, would create a distribution in the nuclear Zeeman temperature.   This effect would compromise the spatial identity of the relaxation profile discussed in this review of HTS.

Diffusion can be attributed to the nuclear dipole-dipole interaction, via a 'flip-flop' process, or similarly by any indirect electronic interaction, as is well known to exist in some metals such as platinum and thallium.  The measurements on vanadium superconducting alloys by Silbernagel \et\cite{sil66,sil67} indicated an extra  relaxation mechanism for \rt\, at low temperatures beyond that expected for an $s$-wave BCS state.  This was qualitatively identified with nuclear  spin diffusion of magnetization between the vortex core and the bulk of the superconductor.  In these experiments a rough estimate was made of the diffusion coefficient from \rttwo\, data. Later Genack and Redfield\cite{gen73,gen75} used field cycling techniques with samples of vanadium metal, exploiting the large gradients that can exist in the mixed state of a type II superconductor.  They measured the diffusion coefficient, D,  and found reasonable agreement with the theory.  But more importantly, they determined that spin diffusion in the superconducting mixed state is thermodynamically quenched in a very short time owing to depletion of the dipole energy reservoir.  In these circumstances diffusion takes place with a diffusion constant, $D_{eff}$ that is reduced by a very substantial factor from its normal value,
\begin{equation} 
D_{eff}\,\,\,= \,\,\,D\,\,( \frac{H_d}{\Delta H})^2
\end{equation}

\noindent
This reduction factor can be of order $\sim 10^{-3}$, and is proportional to the square of the product  of natural abundance and the gyromagnetic ratio.  Here $H_d$ is the dipolar field and $\Delta H$ is the maximal variation in the local field in the mixed state. Consequently, these authors ruled out the diffusion process in the earlier interpretations of the \rt\, experiments.\cite{sil66,sil67}  More recently Wortis\cite{wor98} studied the possible effects of spin diffusion on \rt\, in cuprate superconductors and compared her theoretical calculations with experiment.\cite{cur00} She considered \YBCO\, with realistic experimental conditions, allowing for spin diffusion coefficients determined from a combination of direct dipole-dipole coupling and indirect interactions.  Her conclusions were the same as Genack and Redfield;\cite{gen73,gen75}  $i.e.$ spin diffusion does not play a significant role in HTS spin-lattice relaxation.

\bibstyle{apsrev.bst}
\bibliography{Reviewbib}

\begin{thebibliography}{70}%
\makeatletter
\providecommand \@ifxundefined [1]{%
 \@ifx{#1\undefined}
}%
\providecommand \@ifnum [1]{%
 \ifnum #1\expandafter \@firstoftwo
 \else \expandafter \@secondoftwo
 \fi
}%
\providecommand \@ifx [1]{%
 \ifx #1\expandafter \@firstoftwo
 \else \expandafter \@secondoftwo
 \fi
}%
\providecommand \natexlab [1]{#1}%
\providecommand \enquote  [1]{``#1''}%
\providecommand \bibnamefont  [1]{#1}%
\providecommand \bibfnamefont [1]{#1}%
\providecommand \citenamefont [1]{#1}%
\providecommand \href@noop [0]{\@secondoftwo}%
\providecommand \href [0]{\begingroup \@sanitize@url \@href}%
\providecommand \@href[1]{\@@startlink{#1}\@@href}%
\providecommand \@@href[1]{\endgroup#1\@@endlink}%
\providecommand \@sanitize@url [0]{\catcode `\\12\catcode `\$12\catcode
  `\&12\catcode `\#12\catcode `\^12\catcode `\_12\catcode `\%12\relax}%
\providecommand \@@startlink[1]{}%
\providecommand \@@endlink[0]{}%
\providecommand \url  [0]{\begingroup\@sanitize@url \@url }%
\providecommand \@url [1]{\endgroup\@href {#1}{\urlprefix }}%
\providecommand \urlprefix  [0]{URL }%
\providecommand \Eprint [0]{\href }%
\providecommand \doibase [0]{http://dx.doi.org/}%
\providecommand \selectlanguage [0]{\@gobble}%
\providecommand \bibinfo  [0]{\@secondoftwo}%
\providecommand \bibfield  [0]{\@secondoftwo}%
\providecommand \translation [1]{[#1]}%
\providecommand \BibitemOpen [0]{}%
\providecommand \bibitemStop [0]{}%
\providecommand \bibitemNoStop [0]{.\EOS\space}%
\providecommand \EOS [0]{\spacefactor3000\relax}%
\providecommand \BibitemShut  [1]{\csname bibitem#1\endcsname}%
\let\auto@bib@innerbib\@empty
\bibitem [{\citenamefont {Bednorz}\ and\ \citenamefont
  {M\"uller}(1986)}]{bed86}%
  \BibitemOpen
  \bibfield  {author} {\bibinfo {author} {\bibfnamefont {J.~G.}\ \bibnamefont
  {Bednorz}}\ and\ \bibinfo {author} {\bibfnamefont {K.~A.}\ \bibnamefont
  {M\"uller}},\ }\href@noop {} {\bibfield  {journal} {\bibinfo  {journal} {Z.
  Physik, B}\ }\textbf {\bibinfo {volume} {64}},\ \bibinfo {pages} {189}
  (\bibinfo {year} {1986})}\BibitemShut {NoStop}%
\bibitem [{\citenamefont {Lee}\ \emph {et~al.}(1987)\citenamefont {Lee},
  \citenamefont {Yudkowsky}, \citenamefont {Halperin}, \citenamefont {Thiel},
  \citenamefont {Hwu},\ and\ \citenamefont {Poeppelmeier}}]{lee87}%
  \BibitemOpen
  \bibfield  {author} {\bibinfo {author} {\bibfnamefont {M.}~\bibnamefont
  {Lee}}, \bibinfo {author} {\bibfnamefont {M.}~\bibnamefont {Yudkowsky}},
  \bibinfo {author} {\bibfnamefont {W.~P.}\ \bibnamefont {Halperin}}, \bibinfo
  {author} {\bibfnamefont {J.}~\bibnamefont {Thiel}}, \bibinfo {author}
  {\bibfnamefont {S.~J.}\ \bibnamefont {Hwu}}, \ and\ \bibinfo {author}
  {\bibfnamefont {K.~R.}\ \bibnamefont {Poeppelmeier}},\ }\href {\doibase
  10.1103/PhysRevB.36.2378} {\bibfield  {journal} {\bibinfo  {journal} {Phys.
  Rev. B}\ }\textbf {\bibinfo {volume} {36}},\ \bibinfo {pages} {2378}
  (\bibinfo {year} {1987})}\BibitemShut {NoStop}%
\bibitem [{\citenamefont {Takigawa}\ \emph {et~al.}(1989)\citenamefont
  {Takigawa}, \citenamefont {Hammel}, \citenamefont {Heffner}, \citenamefont
  {Fisk}, \citenamefont {Ott},\ and\ \citenamefont {Thompson}}]{tak89}%
  \BibitemOpen
  \bibfield  {author} {\bibinfo {author} {\bibfnamefont {M.}~\bibnamefont
  {Takigawa}}, \bibinfo {author} {\bibfnamefont {P.~C.}\ \bibnamefont
  {Hammel}}, \bibinfo {author} {\bibfnamefont {R.~H.}\ \bibnamefont {Heffner}},
  \bibinfo {author} {\bibfnamefont {Z.}~\bibnamefont {Fisk}}, \bibinfo {author}
  {\bibfnamefont {K.~C.}\ \bibnamefont {Ott}}, \ and\ \bibinfo {author}
  {\bibfnamefont {J.~D.}\ \bibnamefont {Thompson}},\ }\href {\doibase
  10.1103/PhysRevLett.63.1865} {\bibfield  {journal} {\bibinfo  {journal}
  {Phys. Rev. Lett.}\ }\textbf {\bibinfo {volume} {63}},\ \bibinfo {pages}
  {1865} (\bibinfo {year} {1989})}\BibitemShut {NoStop}%
\bibitem [{\citenamefont {Lee}\ \emph {et~al.}(1989)\citenamefont {Lee},
  \citenamefont {Song}, \citenamefont {Halperin}, \citenamefont {Tonge},
  \citenamefont {Marks}, \citenamefont {Marcy},\ and\ \citenamefont
  {Kannewurf}}]{lee89}%
  \BibitemOpen
  \bibfield  {author} {\bibinfo {author} {\bibfnamefont {M.}~\bibnamefont
  {Lee}}, \bibinfo {author} {\bibfnamefont {Y.-Q.}\ \bibnamefont {Song}},
  \bibinfo {author} {\bibfnamefont {W.~P.}\ \bibnamefont {Halperin}}, \bibinfo
  {author} {\bibfnamefont {L.~M.}\ \bibnamefont {Tonge}}, \bibinfo {author}
  {\bibfnamefont {T.~J.}\ \bibnamefont {Marks}}, \bibinfo {author}
  {\bibfnamefont {H.~O.}\ \bibnamefont {Marcy}}, \ and\ \bibinfo {author}
  {\bibfnamefont {C.~R.}\ \bibnamefont {Kannewurf}},\ }\href {\doibase
  10.1103/PhysRevB.40.817} {\bibfield  {journal} {\bibinfo  {journal} {Phys.
  Rev. B}\ }\textbf {\bibinfo {volume} {40}},\ \bibinfo {pages} {817} (\bibinfo
  {year} {1989})}\BibitemShut {NoStop}%
\bibitem [{\citenamefont {Brandt}(1991)}]{bra91}%
  \BibitemOpen
  \bibfield  {author} {\bibinfo {author} {\bibfnamefont {E.~H.}\ \bibnamefont
  {Brandt}},\ }\href {\doibase 10.1103/PhysRevLett.66.3213} {\bibfield
  {journal} {\bibinfo  {journal} {Phys. Rev. Lett.}\ }\textbf {\bibinfo
  {volume} {66}},\ \bibinfo {pages} {3213} (\bibinfo {year}
  {1991})}\BibitemShut {NoStop}%
\bibitem [{\citenamefont {Mitrovi\'c}(2001)}]{mitthesis}%
  \BibitemOpen
  \bibfield  {author} {\bibinfo {author} {\bibfnamefont {V.~F.}\ \bibnamefont
  {Mitrovi\'c}},\ }\href@noop {} {Ph.D. thesis},\ \bibinfo  {school}
  {Northwestern University} (\bibinfo {year} {2001})\BibitemShut {NoStop}%
\bibitem [{\citenamefont {MacLaughlin}(1976)}]{mac76}%
  \BibitemOpen
  \bibfield  {author} {\bibinfo {author} {\bibfnamefont {D.~E.}\ \bibnamefont
  {MacLaughlin}},\ }\href@noop {} {\emph {\bibinfo {title} {Solid State
  Physics}}}\ (\bibinfo  {publisher} {Academic Press},\ \bibinfo {year}
  {1976})\ pp.\ \bibinfo {pages} {1--69}\BibitemShut {NoStop}%
\bibitem [{\citenamefont {Mitrovi\'c}\ \emph {et~al.}(2001)\citenamefont
  {Mitrovi\'c}, \citenamefont {Sigmund}, \citenamefont {Eschrig}, \citenamefont
  {Bachman}, \citenamefont {Halperin}, \citenamefont {Reyes}, \citenamefont
  {Kuhns},\ and\ \citenamefont {Moulton}}]{mit01b}%
  \BibitemOpen
  \bibfield  {author} {\bibinfo {author} {\bibfnamefont {V.~F.}\ \bibnamefont
  {Mitrovi\'c}}, \bibinfo {author} {\bibfnamefont {E.~E.}\ \bibnamefont
  {Sigmund}}, \bibinfo {author} {\bibfnamefont {M.}~\bibnamefont {Eschrig}},
  \bibinfo {author} {\bibfnamefont {H.~N.}\ \bibnamefont {Bachman}}, \bibinfo
  {author} {\bibfnamefont {W.~P.}\ \bibnamefont {Halperin}}, \bibinfo {author}
  {\bibfnamefont {A.~P.}\ \bibnamefont {Reyes}}, \bibinfo {author}
  {\bibfnamefont {P.}~\bibnamefont {Kuhns}}, \ and\ \bibinfo {author}
  {\bibfnamefont {W.~G.}\ \bibnamefont {Moulton}},\ }\href@noop {} {\bibfield
  {journal} {\bibinfo  {journal} {Nature}\ }\textbf {\bibinfo {volume} {413}},\
  \bibinfo {pages} {501} (\bibinfo {year} {2001})}\BibitemShut {NoStop}%
\bibitem [{\citenamefont {Pennington}\ and\ \citenamefont
  {Shlichter}(1990)}]{pen90}%
  \BibitemOpen
  \bibfield  {author} {\bibinfo {author} {\bibfnamefont {C.~H.}\ \bibnamefont
  {Pennington}}\ and\ \bibinfo {author} {\bibfnamefont {C.~P.}\ \bibnamefont
  {Shlichter}},\ }\href@noop {} {\emph {\bibinfo {title} {Physical Properties
  of High Temperature Superconductors vol II}}},\ edited by\ \bibinfo {editor}
  {\bibfnamefont {D.~M.}\ \bibnamefont {Ginsberg}}\ (\bibinfo  {publisher}
  {Singapore: World Scientific},\ \bibinfo {year} {1990})\ pp.\ \bibinfo
  {pages} {269 -- 367}\BibitemShut {NoStop}%
\bibitem [{\citenamefont {Asayama}\ \emph {et~al.}(1996)\citenamefont
  {Asayama}, \citenamefont {Kitaoka}, \citenamefont {qing Zheng},\ and\
  \citenamefont {Ishida}}]{asa96}%
  \BibitemOpen
  \bibfield  {author} {\bibinfo {author} {\bibfnamefont {K.}~\bibnamefont
  {Asayama}}, \bibinfo {author} {\bibfnamefont {Y.}~\bibnamefont {Kitaoka}},
  \bibinfo {author} {\bibfnamefont {G.}~\bibnamefont {qing Zheng}}, \ and\
  \bibinfo {author} {\bibfnamefont {K.}~\bibnamefont {Ishida}},\ }\href
  {\doibase 10.1016/0079-6565(95)01025-4} {\bibfield  {journal} {\bibinfo
  {journal} {Progress in Nuclear Magnetic Resonance Spectroscopy}\ }\textbf
  {\bibinfo {volume} {28}},\ \bibinfo {pages} {221 } (\bibinfo {year}
  {1996})}\BibitemShut {NoStop}%
\bibitem [{\citenamefont {Berthier}\ \emph {et~al.}(1996)\citenamefont
  {Berthier}, \citenamefont {Julien}, \citenamefont {Horvati\'c},\ and\
  \citenamefont {Berthier}}]{ber96}%
  \BibitemOpen
  \bibfield  {author} {\bibinfo {author} {\bibfnamefont {C.}~\bibnamefont
  {Berthier}}, \bibinfo {author} {\bibfnamefont {M.}~\bibnamefont {Julien}},
  \bibinfo {author} {\bibfnamefont {M.}~\bibnamefont {Horvati\'c}}, \ and\
  \bibinfo {author} {\bibfnamefont {Y.}~\bibnamefont {Berthier}},\ }\href@noop
  {} {\bibfield  {journal} {\bibinfo  {journal} {J. Phys. I}\ }\textbf
  {\bibinfo {volume} {6}},\ \bibinfo {pages} {2205 } (\bibinfo {year}
  {1996})}\BibitemShut {NoStop}%
\bibitem [{\citenamefont {Rigamonti}\ \emph {et~al.}(1998)\citenamefont
  {Rigamonti}, \citenamefont {Borsa},\ and\ \citenamefont {Carrretta}}]{rig98}%
  \BibitemOpen
  \bibfield  {author} {\bibinfo {author} {\bibfnamefont {A.}~\bibnamefont
  {Rigamonti}}, \bibinfo {author} {\bibfnamefont {F.}~\bibnamefont {Borsa}}, \
  and\ \bibinfo {author} {\bibfnamefont {P.}~\bibnamefont {Carrretta}},\
  }\href@noop {} {\bibfield  {journal} {\bibinfo  {journal} {Rep. Prog. Phys.}\
  }\textbf {\bibinfo {volume} {61}},\ \bibinfo {pages} {1367 } (\bibinfo {year}
  {1998})}\BibitemShut {NoStop}%
\bibitem [{\citenamefont {Walstedt}(2008)}]{wal08}%
  \BibitemOpen
  \bibfield  {author} {\bibinfo {author} {\bibfnamefont {R.~E.}\ \bibnamefont
  {Walstedt}},\ }\href@noop {} {\emph {\bibinfo {title} {The NMR Probe of
  High-T$_c$ Materials}}}\ (\bibinfo  {publisher} {Springer},\ \bibinfo {year}
  {2008})\BibitemShut {NoStop}%
\bibitem [{\citenamefont {Curro}(2009)}]{cur09}%
  \BibitemOpen
  \bibfield  {author} {\bibinfo {author} {\bibfnamefont {N.~J.}\ \bibnamefont
  {Curro}},\ }\href@noop {} {\bibfield  {journal} {\bibinfo  {journal} {Rep.
  Prog. Phys.}\ }\textbf {\bibinfo {volume} {72}},\ \bibinfo {pages} {026502}
  (\bibinfo {year} {2009})}\BibitemShut {NoStop}%
\bibitem [{\citenamefont {Abrikosov}(1957)}]{abr57}%
  \BibitemOpen
  \bibfield  {author} {\bibinfo {author} {\bibfnamefont {A.~A.}\ \bibnamefont
  {Abrikosov}},\ }\href@noop {} {\bibfield  {journal} {\bibinfo  {journal}
  {Sov. Phys. JEPT}\ }\textbf {\bibinfo {volume} {5}},\ \bibinfo {pages} {1174}
  (\bibinfo {year} {1957})}\BibitemShut {NoStop}%
\bibitem [{\citenamefont {London}(1950)}]{lon50}%
  \BibitemOpen
  \bibfield  {author} {\bibinfo {author} {\bibfnamefont {F.}~\bibnamefont
  {London}},\ }\href@noop {} {\emph {\bibinfo {title} {Superfluids}}},\
  Vol.~\bibinfo {volume} {1}\ (\bibinfo  {publisher} {Wiley, New York},\
  \bibinfo {year} {1950})\BibitemShut {NoStop}%
\bibitem [{\citenamefont {Mitrovi\'c}\ \emph
  {et~al.}(2003{\natexlab{a}})\citenamefont {Mitrovi\'c}, \citenamefont
  {Sigmund},\ and\ \citenamefont {Halperin}}]{mit03b}%
  \BibitemOpen
  \bibfield  {author} {\bibinfo {author} {\bibfnamefont {V.}~\bibnamefont
  {Mitrovi\'c}}, \bibinfo {author} {\bibfnamefont {E.}~\bibnamefont {Sigmund}},
  \ and\ \bibinfo {author} {\bibfnamefont {W.}~\bibnamefont {Halperin}},\
  }\href@noop {} {\bibfield  {journal} {\bibinfo  {journal} {Physica C:
  Superconductivity}\ }\textbf {\bibinfo {volume} {388-389}},\ \bibinfo {pages}
  {629 } (\bibinfo {year} {2003}{\natexlab{a}})}\BibitemShut {NoStop}%
\bibitem [{\citenamefont {Zheng}\ \emph {et~al.}(2002)\citenamefont {Zheng},
  \citenamefont {Ozaki}, \citenamefont {Kitaoka}, \citenamefont {Kuhns},
  \citenamefont {Reyes},\ and\ \citenamefont {Moulton}}]{zhe02}%
  \BibitemOpen
  \bibfield  {author} {\bibinfo {author} {\bibfnamefont {G.-q.}\ \bibnamefont
  {Zheng}}, \bibinfo {author} {\bibfnamefont {H.}~\bibnamefont {Ozaki}},
  \bibinfo {author} {\bibfnamefont {Y.}~\bibnamefont {Kitaoka}}, \bibinfo
  {author} {\bibfnamefont {P.}~\bibnamefont {Kuhns}}, \bibinfo {author}
  {\bibfnamefont {A.~P.}\ \bibnamefont {Reyes}}, \ and\ \bibinfo {author}
  {\bibfnamefont {W.~G.}\ \bibnamefont {Moulton}},\ }\href@noop {} {\bibfield
  {journal} {\bibinfo  {journal} {Phys. Rev. Lett.}\ }\textbf {\bibinfo
  {volume} {88}},\ \bibinfo {pages} {077003} (\bibinfo {year}
  {2002})}\BibitemShut {NoStop}%
\bibitem [{\citenamefont {Kakuyanagi}\ \emph {et~al.}(2003)\citenamefont
  {Kakuyanagi}, \citenamefont {Kumagai}, \citenamefont {Matsuda},\ and\
  \citenamefont {Hasegawa}}]{kak03}%
  \BibitemOpen
  \bibfield  {author} {\bibinfo {author} {\bibfnamefont {K.}~\bibnamefont
  {Kakuyanagi}}, \bibinfo {author} {\bibfnamefont {K.}~\bibnamefont {Kumagai}},
  \bibinfo {author} {\bibfnamefont {Y.}~\bibnamefont {Matsuda}}, \ and\
  \bibinfo {author} {\bibfnamefont {M.}~\bibnamefont {Hasegawa}},\ }\href
  {\doibase 10.1103/PhysRevLett.90.197003} {\bibfield  {journal} {\bibinfo
  {journal} {Phys. Rev. Lett.}\ }\textbf {\bibinfo {volume} {90}},\ \bibinfo
  {pages} {197003} (\bibinfo {year} {2003})}\BibitemShut {NoStop}%
\bibitem [{\citenamefont {Mounce}\ \emph {et~al.}(2011)\citenamefont {Mounce},
  \citenamefont {Oh}, \citenamefont {Mukhopadhyay}, \citenamefont {Halperin},
  \citenamefont {Reyes}, \citenamefont {Kuhns}, \citenamefont {Fujita},
  \citenamefont {Ishikado},\ and\ \citenamefont {Uchida}}]{mou11b}%
  \BibitemOpen
  \bibfield  {author} {\bibinfo {author} {\bibfnamefont {A.~M.}\ \bibnamefont
  {Mounce}}, \bibinfo {author} {\bibfnamefont {S.}~\bibnamefont {Oh}}, \bibinfo
  {author} {\bibfnamefont {S.}~\bibnamefont {Mukhopadhyay}}, \bibinfo {author}
  {\bibfnamefont {W.~P.}\ \bibnamefont {Halperin}}, \bibinfo {author}
  {\bibfnamefont {A.~P.}\ \bibnamefont {Reyes}}, \bibinfo {author}
  {\bibfnamefont {P.~L.}\ \bibnamefont {Kuhns}}, \bibinfo {author}
  {\bibfnamefont {K.}~\bibnamefont {Fujita}}, \bibinfo {author} {\bibfnamefont
  {M.}~\bibnamefont {Ishikado}}, \ and\ \bibinfo {author} {\bibfnamefont
  {S.}~\bibnamefont {Uchida}},\ }\href {\doibase
  10.1103/PhysRevLett.106.057003} {\bibfield  {journal} {\bibinfo  {journal}
  {Phys. Rev. Lett.}\ }\textbf {\bibinfo {volume} {106}},\ \bibinfo {pages}
  {057003} (\bibinfo {year} {2011})}\BibitemShut {NoStop}%
\bibitem [{\citenamefont {Chen}\ \emph {et~al.}(2006)\citenamefont {Chen},
  \citenamefont {Sengupta}, \citenamefont {Halperin}, \citenamefont {Sigmund},
  \citenamefont {Mitrovi\'c}, \citenamefont {Lee}, \citenamefont {Kang},
  \citenamefont {Mean}, \citenamefont {Kim},\ and\ \citenamefont
  {Cho}}]{che06}%
  \BibitemOpen
  \bibfield  {author} {\bibinfo {author} {\bibfnamefont {B.}~\bibnamefont
  {Chen}}, \bibinfo {author} {\bibfnamefont {P.}~\bibnamefont {Sengupta}},
  \bibinfo {author} {\bibfnamefont {W.~P.}\ \bibnamefont {Halperin}}, \bibinfo
  {author} {\bibfnamefont {E.~E.}\ \bibnamefont {Sigmund}}, \bibinfo {author}
  {\bibfnamefont {V.~F.}\ \bibnamefont {Mitrovi\'c}}, \bibinfo {author}
  {\bibfnamefont {M.~H.}\ \bibnamefont {Lee}}, \bibinfo {author} {\bibfnamefont
  {K.~H.}\ \bibnamefont {Kang}}, \bibinfo {author} {\bibfnamefont {B.~J.}\
  \bibnamefont {Mean}}, \bibinfo {author} {\bibfnamefont {J.~Y.}\ \bibnamefont
  {Kim}}, \ and\ \bibinfo {author} {\bibfnamefont {B.~K.}\ \bibnamefont
  {Cho}},\ }\href@noop {} {\bibfield  {journal} {\bibinfo  {journal} {New
  Journal of Physics}\ }\textbf {\bibinfo {volume} {8}},\ \bibinfo {pages}
  {274} (\bibinfo {year} {2006})}\BibitemShut {NoStop}%
\bibitem [{\citenamefont {Y.-Q.}\ and\ \citenamefont {Song}(1995)}]{son95}%
  \BibitemOpen
  \bibfield  {author} {\bibinfo {author} {\bibnamefont {Y.-Q.}}\ and\ \bibinfo
  {author} {\bibnamefont {Song}},\ }\href@noop {} {\bibfield  {journal}
  {\bibinfo  {journal} {Physica C: Superconductivity}\ }\textbf {\bibinfo
  {volume} {241}},\ \bibinfo {pages} {187 } (\bibinfo {year}
  {1995})}\BibitemShut {NoStop}%
\bibitem [{\citenamefont {Koutroulakis}\ \emph {et~al.}(2008)\citenamefont
  {Koutroulakis}, \citenamefont {Mitrovi\'c}, \citenamefont {Horvati\'c},
  \citenamefont {Berthier}, \citenamefont {Lapertot},\ and\ \citenamefont
  {Flouquet}}]{kou08}%
  \BibitemOpen
  \bibfield  {author} {\bibinfo {author} {\bibfnamefont {G.}~\bibnamefont
  {Koutroulakis}}, \bibinfo {author} {\bibfnamefont {V.~F.}\ \bibnamefont
  {Mitrovi\'c}}, \bibinfo {author} {\bibfnamefont {M.}~\bibnamefont
  {Horvati\'c}}, \bibinfo {author} {\bibfnamefont {C.}~\bibnamefont
  {Berthier}}, \bibinfo {author} {\bibfnamefont {G.}~\bibnamefont {Lapertot}},
  \ and\ \bibinfo {author} {\bibfnamefont {J.}~\bibnamefont {Flouquet}},\
  }\href {\doibase 10.1103/PhysRevLett.101.047004} {\bibfield  {journal}
  {\bibinfo  {journal} {Phys. Rev. Lett.}\ }\textbf {\bibinfo {volume} {101}},\
  \bibinfo {pages} {047004} (\bibinfo {year} {2008})}\BibitemShut {NoStop}%
\bibitem [{\citenamefont {Blatter}\ \emph {et~al.}(1994)\citenamefont
  {Blatter}, \citenamefont {Feigel'man}, \citenamefont {Geshkenbein},
  \citenamefont {Larkin},\ and\ \citenamefont {Vinokur}}]{bla94}%
  \BibitemOpen
  \bibfield  {author} {\bibinfo {author} {\bibfnamefont {G.}~\bibnamefont
  {Blatter}}, \bibinfo {author} {\bibfnamefont {M.~V.}\ \bibnamefont
  {Feigel'man}}, \bibinfo {author} {\bibfnamefont {V.~B.}\ \bibnamefont
  {Geshkenbein}}, \bibinfo {author} {\bibfnamefont {A.~I.}\ \bibnamefont
  {Larkin}}, \ and\ \bibinfo {author} {\bibfnamefont {V.~M.}\ \bibnamefont
  {Vinokur}},\ }\href {\doibase 10.1103/RevModPhys.66.1125} {\bibfield
  {journal} {\bibinfo  {journal} {Rev. Mod. Phys.}\ }\textbf {\bibinfo {volume}
  {66}},\ \bibinfo {pages} {1125} (\bibinfo {year} {1994})}\BibitemShut
  {NoStop}%
\bibitem [{\citenamefont {Kwok}\ \emph {et~al.}(1992)\citenamefont {Kwok},
  \citenamefont {Fleshler}, \citenamefont {Welp}, \citenamefont {Vinokur},
  \citenamefont {Downey}, \citenamefont {Crabtree},\ and\ \citenamefont
  {Miller}}]{kwo92}%
  \BibitemOpen
  \bibfield  {author} {\bibinfo {author} {\bibfnamefont {W.~K.}\ \bibnamefont
  {Kwok}}, \bibinfo {author} {\bibfnamefont {S.}~\bibnamefont {Fleshler}},
  \bibinfo {author} {\bibfnamefont {U.}~\bibnamefont {Welp}}, \bibinfo {author}
  {\bibfnamefont {V.~M.}\ \bibnamefont {Vinokur}}, \bibinfo {author}
  {\bibfnamefont {J.}~\bibnamefont {Downey}}, \bibinfo {author} {\bibfnamefont
  {G.~W.}\ \bibnamefont {Crabtree}}, \ and\ \bibinfo {author} {\bibfnamefont
  {M.~M.}\ \bibnamefont {Miller}},\ }\href {\doibase
  10.1103/PhysRevLett.69.3370} {\bibfield  {journal} {\bibinfo  {journal}
  {Phys. Rev. Lett.}\ }\textbf {\bibinfo {volume} {69}},\ \bibinfo {pages}
  {3370} (\bibinfo {year} {1992})}\BibitemShut {NoStop}%
\bibitem [{\citenamefont {Fuchs}\ \emph {et~al.}(1998)\citenamefont {Fuchs},
  \citenamefont {Zeldov}, \citenamefont {Tamegai}, \citenamefont {Ooi},
  \citenamefont {Rappaport},\ and\ \citenamefont {Shtrikman}}]{fuc98}%
  \BibitemOpen
  \bibfield  {author} {\bibinfo {author} {\bibfnamefont {D.~T.}\ \bibnamefont
  {Fuchs}}, \bibinfo {author} {\bibfnamefont {E.}~\bibnamefont {Zeldov}},
  \bibinfo {author} {\bibfnamefont {T.}~\bibnamefont {Tamegai}}, \bibinfo
  {author} {\bibfnamefont {S.}~\bibnamefont {Ooi}}, \bibinfo {author}
  {\bibfnamefont {M.}~\bibnamefont {Rappaport}}, \ and\ \bibinfo {author}
  {\bibfnamefont {H.}~\bibnamefont {Shtrikman}},\ }\href@noop {} {\bibfield
  {journal} {\bibinfo  {journal} {Phys. Rev. Lett.}\ }\textbf {\bibinfo
  {volume} {80}},\ \bibinfo {pages} {4971} (\bibinfo {year}
  {1998})}\BibitemShut {NoStop}%
\bibitem [{\citenamefont {Chen}\ \emph {et~al.}(2007)\citenamefont {Chen},
  \citenamefont {Halperin}, \citenamefont {Guptasarma}, \citenamefont {Hinks},
  \citenamefont {Mitrovi\'c}, \citenamefont {Reyes},\ and\ \citenamefont
  {Kuhns}}]{che07}%
  \BibitemOpen
  \bibfield  {author} {\bibinfo {author} {\bibfnamefont {B.}~\bibnamefont
  {Chen}}, \bibinfo {author} {\bibfnamefont {W.}~\bibnamefont {Halperin}},
  \bibinfo {author} {\bibfnamefont {P.}~\bibnamefont {Guptasarma}}, \bibinfo
  {author} {\bibfnamefont {D.}~\bibnamefont {Hinks}}, \bibinfo {author}
  {\bibfnamefont {V.~F.}\ \bibnamefont {Mitrovi\'c}}, \bibinfo {author}
  {\bibfnamefont {A.}~\bibnamefont {Reyes}}, \ and\ \bibinfo {author}
  {\bibfnamefont {P.}~\bibnamefont {Kuhns}},\ }\href {\doibase
  10.1038/nphys540} {\bibfield  {journal} {\bibinfo  {journal} {Nature
  Physics}\ }\textbf {\bibinfo {volume} {3}},\ \bibinfo {pages} {239} (\bibinfo
  {year} {2007})}\BibitemShut {NoStop}%
\bibitem [{\citenamefont {Bachman}\ \emph {et~al.}(1998)\citenamefont
  {Bachman}, \citenamefont {Reyes}, \citenamefont {Mitrovi\'c}, \citenamefont
  {Halperin}, \citenamefont {Kleinhammes}, \citenamefont {Kuhns},\ and\
  \citenamefont {Moulton}}]{bac98}%
  \BibitemOpen
  \bibfield  {author} {\bibinfo {author} {\bibfnamefont {H.~N.}\ \bibnamefont
  {Bachman}}, \bibinfo {author} {\bibfnamefont {A.~P.}\ \bibnamefont {Reyes}},
  \bibinfo {author} {\bibfnamefont {V.~F.}\ \bibnamefont {Mitrovi\'c}},
  \bibinfo {author} {\bibfnamefont {W.~P.}\ \bibnamefont {Halperin}}, \bibinfo
  {author} {\bibfnamefont {A.}~\bibnamefont {Kleinhammes}}, \bibinfo {author}
  {\bibfnamefont {P.}~\bibnamefont {Kuhns}}, \ and\ \bibinfo {author}
  {\bibfnamefont {W.~G.}\ \bibnamefont {Moulton}},\ }\href {\doibase
  10.1103/PhysRevLett.80.1726} {\bibfield  {journal} {\bibinfo  {journal}
  {Phys. Rev. Lett.}\ }\textbf {\bibinfo {volume} {80}},\ \bibinfo {pages}
  {1726} (\bibinfo {year} {1998})}\BibitemShut {NoStop}%
\bibitem [{\citenamefont {Oh}\ \emph {et~al.}(2011{\natexlab{a}})\citenamefont
  {Oh}, \citenamefont {Mounce}, \citenamefont {Mukhopadhyay}, \citenamefont
  {Halperin}, \citenamefont {Vorontsov}, \citenamefont {Bud'ko}, \citenamefont
  {Canfield}, \citenamefont {Furukawa}, \citenamefont {Reyes},\ and\
  \citenamefont {Kuhns}}]{oh11a}%
  \BibitemOpen
  \bibfield  {author} {\bibinfo {author} {\bibfnamefont {S.}~\bibnamefont
  {Oh}}, \bibinfo {author} {\bibfnamefont {A.~M.}\ \bibnamefont {Mounce}},
  \bibinfo {author} {\bibfnamefont {S.}~\bibnamefont {Mukhopadhyay}}, \bibinfo
  {author} {\bibfnamefont {W.~P.}\ \bibnamefont {Halperin}}, \bibinfo {author}
  {\bibfnamefont {A.~B.}\ \bibnamefont {Vorontsov}}, \bibinfo {author}
  {\bibfnamefont {S.~L.}\ \bibnamefont {Bud'ko}}, \bibinfo {author}
  {\bibfnamefont {P.~C.}\ \bibnamefont {Canfield}}, \bibinfo {author}
  {\bibfnamefont {Y.}~\bibnamefont {Furukawa}}, \bibinfo {author}
  {\bibfnamefont {A.~P.}\ \bibnamefont {Reyes}}, \ and\ \bibinfo {author}
  {\bibfnamefont {P.~L.}\ \bibnamefont {Kuhns}},\ }\href@noop {} {\bibfield
  {journal} {\bibinfo  {journal} {Phys. Rev. B}\ }\textbf {\bibinfo {volume}
  {83}},\ \bibinfo {pages} {214501} (\bibinfo {year}
  {2011}{\natexlab{a}})}\BibitemShut {NoStop}%
\bibitem [{\citenamefont {Reyes}\ \emph {et~al.}(1997)\citenamefont {Reyes},
  \citenamefont {Tang}, \citenamefont {Bachman}, \citenamefont {Halperin},
  \citenamefont {Martindale},\ and\ \citenamefont {Hammel}}]{rey97}%
  \BibitemOpen
  \bibfield  {author} {\bibinfo {author} {\bibfnamefont {A.~P.}\ \bibnamefont
  {Reyes}}, \bibinfo {author} {\bibfnamefont {X.~P.}\ \bibnamefont {Tang}},
  \bibinfo {author} {\bibfnamefont {H.~N.}\ \bibnamefont {Bachman}}, \bibinfo
  {author} {\bibfnamefont {W.~P.}\ \bibnamefont {Halperin}}, \bibinfo {author}
  {\bibfnamefont {J.~A.}\ \bibnamefont {Martindale}}, \ and\ \bibinfo {author}
  {\bibfnamefont {P.~C.}\ \bibnamefont {Hammel}},\ }\href {\doibase
  10.1103/PhysRevB.55.R14737} {\bibfield  {journal} {\bibinfo  {journal} {Phys.
  Rev. B}\ }\textbf {\bibinfo {volume} {55}},\ \bibinfo {pages} {R14737}
  (\bibinfo {year} {1997})}\BibitemShut {NoStop}%
\bibitem [{\citenamefont {Glazman}\ and\ \citenamefont
  {Koshelev}(1991)}]{gla91}%
  \BibitemOpen
  \bibfield  {author} {\bibinfo {author} {\bibfnamefont {L.~I.}\ \bibnamefont
  {Glazman}}\ and\ \bibinfo {author} {\bibfnamefont {A.~E.}\ \bibnamefont
  {Koshelev}},\ }\href {\doibase 10.1103/PhysRevB.43.2835} {\bibfield
  {journal} {\bibinfo  {journal} {Phys. Rev. B}\ }\textbf {\bibinfo {volume}
  {43}},\ \bibinfo {pages} {2835} (\bibinfo {year} {1991})}\BibitemShut
  {NoStop}%
\bibitem [{\citenamefont {Yip}\ and\ \citenamefont {Sauls}(1992)}]{yip92}%
  \BibitemOpen
  \bibfield  {author} {\bibinfo {author} {\bibfnamefont {S.~K.}\ \bibnamefont
  {Yip}}\ and\ \bibinfo {author} {\bibfnamefont {J.~A.}\ \bibnamefont
  {Sauls}},\ }\href {\doibase 10.1103/PhysRevLett.69.2264} {\bibfield
  {journal} {\bibinfo  {journal} {Phys. Rev. Lett.}\ }\textbf {\bibinfo
  {volume} {69}},\ \bibinfo {pages} {2264} (\bibinfo {year}
  {1992})}\BibitemShut {NoStop}%
\bibitem [{\citenamefont {Volovik}(1993)}]{vol93}%
  \BibitemOpen
  \bibfield  {author} {\bibinfo {author} {\bibfnamefont {G.~E.}\ \bibnamefont
  {Volovik}},\ }\href@noop {} {\bibfield  {journal} {\bibinfo  {journal}
  {Pis'ma Zh. Eksp. Teor. Fiz.}\ }\textbf {\bibinfo {volume} {58}} (\bibinfo
  {year} {1993})}\BibitemShut {NoStop}%
\bibitem [{\citenamefont {Moler}\ \emph {et~al.}(1994)\citenamefont {Moler},
  \citenamefont {Baar}, \citenamefont {Urbach}, \citenamefont {Liang},
  \citenamefont {Hardy},\ and\ \citenamefont {Kapitulnik}}]{mol94}%
  \BibitemOpen
  \bibfield  {author} {\bibinfo {author} {\bibfnamefont {K.~A.}\ \bibnamefont
  {Moler}}, \bibinfo {author} {\bibfnamefont {D.~J.}\ \bibnamefont {Baar}},
  \bibinfo {author} {\bibfnamefont {J.~S.}\ \bibnamefont {Urbach}}, \bibinfo
  {author} {\bibfnamefont {R.}~\bibnamefont {Liang}}, \bibinfo {author}
  {\bibfnamefont {W.~N.}\ \bibnamefont {Hardy}}, \ and\ \bibinfo {author}
  {\bibfnamefont {A.}~\bibnamefont {Kapitulnik}},\ }\href@noop {} {\bibfield
  {journal} {\bibinfo  {journal} {Phys. Rev. Lett.}\ }\textbf {\bibinfo
  {volume} {73}},\ \bibinfo {pages} {2744} (\bibinfo {year}
  {1994})}\BibitemShut {NoStop}%
\bibitem [{\citenamefont {Moler}\ \emph {et~al.}(1997)\citenamefont {Moler},
  \citenamefont {Sisson}, \citenamefont {Urbach}, \citenamefont {Beasley},
  \citenamefont {Kapitulnik}, \citenamefont {Baar}, \citenamefont {Liang},\
  and\ \citenamefont {Hardy}}]{mol97}%
  \BibitemOpen
  \bibfield  {author} {\bibinfo {author} {\bibfnamefont {K.~A.}\ \bibnamefont
  {Moler}}, \bibinfo {author} {\bibfnamefont {D.~L.}\ \bibnamefont {Sisson}},
  \bibinfo {author} {\bibfnamefont {J.~S.}\ \bibnamefont {Urbach}}, \bibinfo
  {author} {\bibfnamefont {M.~R.}\ \bibnamefont {Beasley}}, \bibinfo {author}
  {\bibfnamefont {A.}~\bibnamefont {Kapitulnik}}, \bibinfo {author}
  {\bibfnamefont {D.~J.}\ \bibnamefont {Baar}}, \bibinfo {author}
  {\bibfnamefont {R.}~\bibnamefont {Liang}}, \ and\ \bibinfo {author}
  {\bibfnamefont {W.~N.}\ \bibnamefont {Hardy}},\ }\href@noop {} {\bibfield
  {journal} {\bibinfo  {journal} {Phys. Rev. B}\ }\textbf {\bibinfo {volume}
  {55}},\ \bibinfo {pages} {3954} (\bibinfo {year} {1997})}\BibitemShut
  {NoStop}%
\bibitem [{\citenamefont {Aubin}\ \emph {et~al.}(1999)\citenamefont {Aubin},
  \citenamefont {Behnia}, \citenamefont {Ooi},\ and\ \citenamefont
  {Tamegai}}]{aub99}%
  \BibitemOpen
  \bibfield  {author} {\bibinfo {author} {\bibfnamefont {H.}~\bibnamefont
  {Aubin}}, \bibinfo {author} {\bibfnamefont {K.}~\bibnamefont {Behnia}},
  \bibinfo {author} {\bibfnamefont {S.}~\bibnamefont {Ooi}}, \ and\ \bibinfo
  {author} {\bibfnamefont {T.}~\bibnamefont {Tamegai}},\ }\href {\doibase
  10.1103/PhysRevLett.82.624} {\bibfield  {journal} {\bibinfo  {journal} {Phys.
  Rev. Lett.}\ }\textbf {\bibinfo {volume} {82}},\ \bibinfo {pages} {624}
  (\bibinfo {year} {1999})}\BibitemShut {NoStop}%
\bibitem [{\citenamefont {Chiao}\ \emph {et~al.}(2000)\citenamefont {Chiao},
  \citenamefont {Hill}, \citenamefont {Lupien}, \citenamefont {Taillefer},
  \citenamefont {Lambert}, \citenamefont {Gagnon},\ and\ \citenamefont
  {Fournier}}]{chi00}%
  \BibitemOpen
  \bibfield  {author} {\bibinfo {author} {\bibfnamefont {M.}~\bibnamefont
  {Chiao}}, \bibinfo {author} {\bibfnamefont {R.~W.}\ \bibnamefont {Hill}},
  \bibinfo {author} {\bibfnamefont {C.}~\bibnamefont {Lupien}}, \bibinfo
  {author} {\bibfnamefont {L.}~\bibnamefont {Taillefer}}, \bibinfo {author}
  {\bibfnamefont {P.}~\bibnamefont {Lambert}}, \bibinfo {author} {\bibfnamefont
  {R.}~\bibnamefont {Gagnon}}, \ and\ \bibinfo {author} {\bibfnamefont
  {P.}~\bibnamefont {Fournier}},\ }\href {\doibase 10.1103/PhysRevB.62.3554}
  {\bibfield  {journal} {\bibinfo  {journal} {Phys. Rev. B}\ }\textbf {\bibinfo
  {volume} {62}},\ \bibinfo {pages} {3554} (\bibinfo {year}
  {2000})}\BibitemShut {NoStop}%
\bibitem [{\citenamefont {Takigawa}\ \emph {et~al.}(1999)\citenamefont
  {Takigawa}, \citenamefont {Ichioka},\ and\ \citenamefont {Machida}}]{tak99}%
  \BibitemOpen
  \bibfield  {author} {\bibinfo {author} {\bibfnamefont {M.}~\bibnamefont
  {Takigawa}}, \bibinfo {author} {\bibfnamefont {M.}~\bibnamefont {Ichioka}}, \
  and\ \bibinfo {author} {\bibfnamefont {K.}~\bibnamefont {Machida}},\ }\href
  {\doibase 10.1103/PhysRevLett.83.3057} {\bibfield  {journal} {\bibinfo
  {journal} {Phys. Rev. Lett.}\ }\textbf {\bibinfo {volume} {83}},\ \bibinfo
  {pages} {3057} (\bibinfo {year} {1999})}\BibitemShut {NoStop}%
\bibitem [{\citenamefont {Curro}\ \emph {et~al.}(2000)\citenamefont {Curro},
  \citenamefont {Milling}, \citenamefont {Haase},\ and\ \citenamefont
  {Slichter}}]{cur00}%
  \BibitemOpen
  \bibfield  {author} {\bibinfo {author} {\bibfnamefont {N.~J.}\ \bibnamefont
  {Curro}}, \bibinfo {author} {\bibfnamefont {C.}~\bibnamefont {Milling}},
  \bibinfo {author} {\bibfnamefont {J.}~\bibnamefont {Haase}}, \ and\ \bibinfo
  {author} {\bibfnamefont {C.~P.}\ \bibnamefont {Slichter}},\ }\href {\doibase
  10.1103/PhysRevB.62.3473} {\bibfield  {journal} {\bibinfo  {journal} {Phys.
  Rev. B}\ }\textbf {\bibinfo {volume} {62}},\ \bibinfo {pages} {3473}
  (\bibinfo {year} {2000})}\BibitemShut {NoStop}%
\bibitem [{\citenamefont {Haase}\ \emph {et~al.}(1998)\citenamefont {Haase},
  \citenamefont {Curro}, \citenamefont {Stern},\ and\ \citenamefont
  {Slichter}}]{haa98}%
  \BibitemOpen
  \bibfield  {author} {\bibinfo {author} {\bibfnamefont {J.}~\bibnamefont
  {Haase}}, \bibinfo {author} {\bibfnamefont {N.~J.}\ \bibnamefont {Curro}},
  \bibinfo {author} {\bibfnamefont {R.}~\bibnamefont {Stern}}, \ and\ \bibinfo
  {author} {\bibfnamefont {C.~P.}\ \bibnamefont {Slichter}},\ }\href {\doibase
  10.1103/PhysRevLett.81.1489} {\bibfield  {journal} {\bibinfo  {journal}
  {Phys. Rev. Lett.}\ }\textbf {\bibinfo {volume} {81}},\ \bibinfo {pages}
  {1489} (\bibinfo {year} {1998})}\BibitemShut {NoStop}%
\bibitem [{\citenamefont {Wortis}\ \emph {et~al.}(2000)\citenamefont {Wortis},
  \citenamefont {Berlinsky},\ and\ \citenamefont {Kallin}}]{wor00}%
  \BibitemOpen
  \bibfield  {author} {\bibinfo {author} {\bibfnamefont {R.}~\bibnamefont
  {Wortis}}, \bibinfo {author} {\bibfnamefont {A.~J.}\ \bibnamefont
  {Berlinsky}}, \ and\ \bibinfo {author} {\bibfnamefont {C.}~\bibnamefont
  {Kallin}},\ }\href {\doibase 10.1103/PhysRevB.61.12342} {\bibfield  {journal}
  {\bibinfo  {journal} {Phys. Rev. B}\ }\textbf {\bibinfo {volume} {61}},\
  \bibinfo {pages} {12342} (\bibinfo {year} {2000})}\BibitemShut {NoStop}%
\bibitem [{\citenamefont {Kakuyanagi}\ \emph {et~al.}(2002)\citenamefont
  {Kakuyanagi}, \citenamefont {ichi Kumagai},\ and\ \citenamefont
  {Matsuda}}]{kak02}%
  \BibitemOpen
  \bibfield  {author} {\bibinfo {author} {\bibfnamefont {K.}~\bibnamefont
  {Kakuyanagi}}, \bibinfo {author} {\bibfnamefont {K.}~\bibnamefont {ichi
  Kumagai}}, \ and\ \bibinfo {author} {\bibfnamefont {Y.}~\bibnamefont
  {Matsuda}},\ }\href {\doibase DOI: 10.1016/S0022-3697(02)00224-X} {\bibfield
  {journal} {\bibinfo  {journal} {Journal of Physics and Chemistry of Solids}\
  }\textbf {\bibinfo {volume} {63}},\ \bibinfo {pages} {2305 } (\bibinfo {year}
  {2002})}\BibitemShut {NoStop}%
\bibitem [{\citenamefont {Mitrovi\'c}\ \emph
  {et~al.}(2003{\natexlab{b}})\citenamefont {Mitrovi\'c}, \citenamefont
  {Sigmund}, \citenamefont {Halperin}, \citenamefont {Reyes}, \citenamefont
  {Kuhns},\ and\ \citenamefont {Moulton}}]{mit03}%
  \BibitemOpen
  \bibfield  {author} {\bibinfo {author} {\bibfnamefont {V.~F.}\ \bibnamefont
  {Mitrovi\'c}}, \bibinfo {author} {\bibfnamefont {E.~E.}\ \bibnamefont
  {Sigmund}}, \bibinfo {author} {\bibfnamefont {W.~P.}\ \bibnamefont
  {Halperin}}, \bibinfo {author} {\bibfnamefont {A.~P.}\ \bibnamefont {Reyes}},
  \bibinfo {author} {\bibfnamefont {P.}~\bibnamefont {Kuhns}}, \ and\ \bibinfo
  {author} {\bibfnamefont {W.~G.}\ \bibnamefont {Moulton}},\ }\href {\doibase
  10.1103/PhysRevB.67.220503} {\bibfield  {journal} {\bibinfo  {journal} {Phys.
  Rev. B}\ }\textbf {\bibinfo {volume} {67}},\ \bibinfo {pages} {220503}
  (\bibinfo {year} {2003}{\natexlab{b}})}\BibitemShut {NoStop}%
\bibitem [{\citenamefont {Oh}\ \emph {et~al.}(2011{\natexlab{b}})\citenamefont
  {Oh}, \citenamefont {Mounce}, \citenamefont {Mukhopadhyay}, \citenamefont
  {Halperin}, \citenamefont {Vorontsov}, \citenamefont {Bud'ko}, \citenamefont
  {Canfield}, \citenamefont {Furukawa}, \citenamefont {Reyes},\ and\
  \citenamefont {Kuhns}}]{oh11b}%
  \BibitemOpen
  \bibfield  {author} {\bibinfo {author} {\bibfnamefont {S.}~\bibnamefont
  {Oh}}, \bibinfo {author} {\bibfnamefont {A.~M.}\ \bibnamefont {Mounce}},
  \bibinfo {author} {\bibfnamefont {S.}~\bibnamefont {Mukhopadhyay}}, \bibinfo
  {author} {\bibfnamefont {W.~P.}\ \bibnamefont {Halperin}}, \bibinfo {author}
  {\bibfnamefont {A.~B.}\ \bibnamefont {Vorontsov}}, \bibinfo {author}
  {\bibfnamefont {S.~L.}\ \bibnamefont {Bud'ko}}, \bibinfo {author}
  {\bibfnamefont {P.~C.}\ \bibnamefont {Canfield}}, \bibinfo {author}
  {\bibfnamefont {Y.}~\bibnamefont {Furukawa}}, \bibinfo {author}
  {\bibfnamefont {A.~P.}\ \bibnamefont {Reyes}}, \ and\ \bibinfo {author}
  {\bibfnamefont {P.~L.}\ \bibnamefont {Kuhns}},\ }\href@noop {} {\bibfield
  {journal} {\bibinfo  {journal} {{arXiv}:1109.3834v2}\ } (\bibinfo {year}
  {2011}{\natexlab{b}})}\BibitemShut {NoStop}%
\bibitem [{\citenamefont {Morr}(2001)}]{mor01}%
  \BibitemOpen
  \bibfield  {author} {\bibinfo {author} {\bibfnamefont {D.~K.}\ \bibnamefont
  {Morr}},\ }\href {\doibase 10.1103/PhysRevB.63.214509} {\bibfield  {journal}
  {\bibinfo  {journal} {Phys. Rev. B}\ }\textbf {\bibinfo {volume} {63}},\
  \bibinfo {pages} {214509} (\bibinfo {year} {2001})}\BibitemShut {NoStop}%
\bibitem [{\citenamefont {Knapp}\ \emph {et~al.}(2002)\citenamefont {Knapp},
  \citenamefont {Kallin}, \citenamefont {Berlinsky},\ and\ \citenamefont
  {Wortis}}]{kna02}%
  \BibitemOpen
  \bibfield  {author} {\bibinfo {author} {\bibfnamefont {D.}~\bibnamefont
  {Knapp}}, \bibinfo {author} {\bibfnamefont {C.}~\bibnamefont {Kallin}},
  \bibinfo {author} {\bibfnamefont {A.~J.}\ \bibnamefont {Berlinsky}}, \ and\
  \bibinfo {author} {\bibfnamefont {R.}~\bibnamefont {Wortis}},\ }\href
  {\doibase 10.1103/PhysRevB.66.144508} {\bibfield  {journal} {\bibinfo
  {journal} {Phys. Rev. B}\ }\textbf {\bibinfo {volume} {66}},\ \bibinfo
  {pages} {144508} (\bibinfo {year} {2002})}\BibitemShut {NoStop}%
\bibitem [{\citenamefont {Throckmorton}\ and\ \citenamefont
  {Vafek}(2010)}]{thr10}%
  \BibitemOpen
  \bibfield  {author} {\bibinfo {author} {\bibfnamefont {R.~E.}\ \bibnamefont
  {Throckmorton}}\ and\ \bibinfo {author} {\bibfnamefont {O.}~\bibnamefont
  {Vafek}},\ }\href@noop {} {\bibfield  {journal} {\bibinfo  {journal} {Phys.
  Rev. B}\ }\textbf {\bibinfo {volume} {81}},\ \bibinfo {pages} {104515}
  (\bibinfo {year} {2010})}\BibitemShut {NoStop}%
\bibitem [{\citenamefont {Recchia}\ \emph {et~al.}(1997)\citenamefont
  {Recchia}, \citenamefont {Martindale}, \citenamefont {Pennington},
  \citenamefont {Hults},\ and\ \citenamefont {Smith}}]{rec97}%
  \BibitemOpen
  \bibfield  {author} {\bibinfo {author} {\bibfnamefont {C.~H.}\ \bibnamefont
  {Recchia}}, \bibinfo {author} {\bibfnamefont {J.~A.}\ \bibnamefont
  {Martindale}}, \bibinfo {author} {\bibfnamefont {C.~H.}\ \bibnamefont
  {Pennington}}, \bibinfo {author} {\bibfnamefont {W.~L.}\ \bibnamefont
  {Hults}}, \ and\ \bibinfo {author} {\bibfnamefont {J.~L.}\ \bibnamefont
  {Smith}},\ }\href {\doibase 10.1103/PhysRevLett.78.3543} {\bibfield
  {journal} {\bibinfo  {journal} {Phys. Rev. Lett.}\ }\textbf {\bibinfo
  {volume} {78}},\ \bibinfo {pages} {3543} (\bibinfo {year}
  {1997})}\BibitemShut {NoStop}%
\bibitem [{\citenamefont {Lu}\ and\ \citenamefont {Wortis}(2006)}]{lu06}%
  \BibitemOpen
  \bibfield  {author} {\bibinfo {author} {\bibfnamefont {T.}~\bibnamefont
  {Lu}}\ and\ \bibinfo {author} {\bibfnamefont {R.}~\bibnamefont {Wortis}},\
  }\href@noop {} {\bibfield  {journal} {\bibinfo  {journal} {Phys. Rev. B}\
  }\textbf {\bibinfo {volume} {74}},\ \bibinfo {eid} {134516} (\bibinfo {year}
  {2006})}\BibitemShut {NoStop}%
\bibitem [{\citenamefont {Walstedt}\ and\ \citenamefont
  {Cheong}(1995)}]{wal95}%
  \BibitemOpen
  \bibfield  {author} {\bibinfo {author} {\bibfnamefont {R.~E.}\ \bibnamefont
  {Walstedt}}\ and\ \bibinfo {author} {\bibfnamefont {S.-W.}\ \bibnamefont
  {Cheong}},\ }\href {\doibase 10.1103/PhysRevB.51.3163} {\bibfield  {journal}
  {\bibinfo  {journal} {Phys. Rev. B}\ }\textbf {\bibinfo {volume} {51}},\
  \bibinfo {pages} {3163} (\bibinfo {year} {1995})}\BibitemShut {NoStop}%
\bibitem [{\citenamefont {Hoffman}\ \emph {et~al.}(2002)\citenamefont
  {Hoffman}, \citenamefont {Hudson}, \citenamefont {Lang}, \citenamefont
  {Madhavan}, \citenamefont {Eisaki}, \citenamefont {Uchida},\ and\
  \citenamefont {Davis}}]{hof02}%
  \BibitemOpen
  \bibfield  {author} {\bibinfo {author} {\bibfnamefont {J.~E.}\ \bibnamefont
  {Hoffman}}, \bibinfo {author} {\bibfnamefont {E.~W.}\ \bibnamefont {Hudson}},
  \bibinfo {author} {\bibfnamefont {K.~M.}\ \bibnamefont {Lang}}, \bibinfo
  {author} {\bibfnamefont {V.}~\bibnamefont {Madhavan}}, \bibinfo {author}
  {\bibfnamefont {H.}~\bibnamefont {Eisaki}}, \bibinfo {author} {\bibfnamefont
  {S.}~\bibnamefont {Uchida}}, \ and\ \bibinfo {author} {\bibfnamefont {J.~C.}\
  \bibnamefont {Davis}},\ }\href {\doibase 10.1126/science.1066974} {\bibfield
  {journal} {\bibinfo  {journal} {Science}\ }\textbf {\bibinfo {volume}
  {295}},\ \bibinfo {pages} {466} (\bibinfo {year} {2002})}\BibitemShut
  {NoStop}%
\bibitem [{\citenamefont {Hanaguri}\ \emph {et~al.}(2004)\citenamefont
  {Hanaguri}, \citenamefont {Lupien}, \citenamefont {Kohsaka}, \citenamefont
  {Lee}, \citenamefont {Azuma}, \citenamefont {Takano}, \citenamefont
  {Takagi},\ and\ \citenamefont {Davis}}]{han04}%
  \BibitemOpen
  \bibfield  {author} {\bibinfo {author} {\bibfnamefont {T.}~\bibnamefont
  {Hanaguri}}, \bibinfo {author} {\bibfnamefont {C.}~\bibnamefont {Lupien}},
  \bibinfo {author} {\bibfnamefont {Y.}~\bibnamefont {Kohsaka}}, \bibinfo
  {author} {\bibfnamefont {D.-H.}\ \bibnamefont {Lee}}, \bibinfo {author}
  {\bibfnamefont {M.}~\bibnamefont {Azuma}}, \bibinfo {author} {\bibfnamefont
  {M.}~\bibnamefont {Takano}}, \bibinfo {author} {\bibfnamefont
  {H.}~\bibnamefont {Takagi}}, \ and\ \bibinfo {author} {\bibfnamefont {J.~C.}\
  \bibnamefont {Davis}},\ }\href@noop {} {\bibfield  {journal} {\bibinfo
  {journal} {Nature}\ }\textbf {\bibinfo {volume} {430}},\ \bibinfo {pages}
  {1001} (\bibinfo {year} {2004})}\BibitemShut {NoStop}%
\bibitem [{\citenamefont {Wise}\ \emph {et~al.}(2008)\citenamefont {Wise},
  \citenamefont {Boyer}, \citenamefont {Chatterjee}, \citenamefont {Kondo},
  \citenamefont {Takeuchi}, \citenamefont {Ikuta}, \citenamefont {Wang},\ and\
  \citenamefont {Hudson}}]{wis08}%
  \BibitemOpen
  \bibfield  {author} {\bibinfo {author} {\bibfnamefont {W.~D.}\ \bibnamefont
  {Wise}}, \bibinfo {author} {\bibfnamefont {M.~C.}\ \bibnamefont {Boyer}},
  \bibinfo {author} {\bibfnamefont {K.}~\bibnamefont {Chatterjee}}, \bibinfo
  {author} {\bibfnamefont {T.}~\bibnamefont {Kondo}}, \bibinfo {author}
  {\bibfnamefont {T.}~\bibnamefont {Takeuchi}}, \bibinfo {author}
  {\bibfnamefont {H.}~\bibnamefont {Ikuta}}, \bibinfo {author} {\bibfnamefont
  {Y.}~\bibnamefont {Wang}}, \ and\ \bibinfo {author} {\bibfnamefont {E.~W.}\
  \bibnamefont {Hudson}},\ }\href@noop {} {\bibfield  {journal} {\bibinfo
  {journal} {Nature Physics}\ }\textbf {\bibinfo {volume} {4}},\ \bibinfo
  {pages} {696} (\bibinfo {year} {2008})}\BibitemShut {NoStop}%
\bibitem [{\citenamefont {Lake}\ \emph {et~al.}(2001)\citenamefont {Lake},
  \citenamefont {Aeppli}, \citenamefont {Clausen}, \citenamefont {McMorrow},
  \citenamefont {Lefmann}, \citenamefont {Hussey}, \citenamefont
  {Mangkorntong}, \citenamefont {Nohara}, \citenamefont {Takagi}, \citenamefont
  {Mason},\ and\ \citenamefont {Schroder}}]{lak01}%
  \BibitemOpen
  \bibfield  {author} {\bibinfo {author} {\bibfnamefont {B.}~\bibnamefont
  {Lake}}, \bibinfo {author} {\bibfnamefont {G.}~\bibnamefont {Aeppli}},
  \bibinfo {author} {\bibfnamefont {K.~N.}\ \bibnamefont {Clausen}}, \bibinfo
  {author} {\bibfnamefont {D.~F.}\ \bibnamefont {McMorrow}}, \bibinfo {author}
  {\bibfnamefont {K.}~\bibnamefont {Lefmann}}, \bibinfo {author} {\bibfnamefont
  {N.~E.}\ \bibnamefont {Hussey}}, \bibinfo {author} {\bibfnamefont
  {N.}~\bibnamefont {Mangkorntong}}, \bibinfo {author} {\bibfnamefont
  {M.}~\bibnamefont {Nohara}}, \bibinfo {author} {\bibfnamefont
  {H.}~\bibnamefont {Takagi}}, \bibinfo {author} {\bibfnamefont {T.~E.}\
  \bibnamefont {Mason}}, \ and\ \bibinfo {author} {\bibfnamefont
  {A.}~\bibnamefont {Schroder}},\ }\href {\doibase 10.1126/science.1056986}
  {\bibfield  {journal} {\bibinfo  {journal} {Science}\ }\textbf {\bibinfo
  {volume} {291}},\ \bibinfo {pages} {1759} (\bibinfo {year}
  {2001})}\BibitemShut {NoStop}%
\bibitem [{\citenamefont {Lake}\ \emph {et~al.}(2002)\citenamefont {Lake},
  \citenamefont {Ronnow}, \citenamefont {Christensen}, \citenamefont {Aeppli},
  \citenamefont {lefmann}, \citenamefont {McMorrow}, \citenamefont
  {Vorderwisch}, \citenamefont {Smeibidl}, \citenamefont {Mangkorntogn},
  \citenamefont {Sasagawa}, \citenamefont {Nohara}, \citenamefont {Takagi},\
  and\ \citenamefont {Mason}}]{lak02}%
  \BibitemOpen
  \bibfield  {author} {\bibinfo {author} {\bibfnamefont {B.}~\bibnamefont
  {Lake}}, \bibinfo {author} {\bibfnamefont {H.}~\bibnamefont {Ronnow}},
  \bibinfo {author} {\bibfnamefont {N.}~\bibnamefont {Christensen}}, \bibinfo
  {author} {\bibfnamefont {G.}~\bibnamefont {Aeppli}}, \bibinfo {author}
  {\bibfnamefont {K.}~\bibnamefont {lefmann}}, \bibinfo {author} {\bibfnamefont
  {D.}~\bibnamefont {McMorrow}}, \bibinfo {author} {\bibfnamefont
  {P.}~\bibnamefont {Vorderwisch}}, \bibinfo {author} {\bibfnamefont
  {P.}~\bibnamefont {Smeibidl}}, \bibinfo {author} {\bibfnamefont
  {N.}~\bibnamefont {Mangkorntogn}}, \bibinfo {author} {\bibfnamefont
  {T.}~\bibnamefont {Sasagawa}}, \bibinfo {author} {\bibfnamefont
  {M.}~\bibnamefont {Nohara}}, \bibinfo {author} {\bibfnamefont
  {H.}~\bibnamefont {Takagi}}, \ and\ \bibinfo {author} {\bibfnamefont
  {T.}~\bibnamefont {Mason}},\ }\href@noop {} {\bibfield  {journal} {\bibinfo
  {journal} {Nature}\ }\textbf {\bibinfo {volume} {415}},\ \bibinfo {pages}
  {299} (\bibinfo {year} {2002})}\BibitemShut {NoStop}%
\bibitem [{\citenamefont {Khaykovich}\ \emph {et~al.}(2002)\citenamefont
  {Khaykovich}, \citenamefont {Lee}, \citenamefont {Erwin}, \citenamefont
  {Lee}, \citenamefont {Wakimoto}, \citenamefont {Thomas}, \citenamefont
  {Kastner},\ and\ \citenamefont {Birgeneau}}]{kha02}%
  \BibitemOpen
  \bibfield  {author} {\bibinfo {author} {\bibfnamefont {B.}~\bibnamefont
  {Khaykovich}}, \bibinfo {author} {\bibfnamefont {Y.~S.}\ \bibnamefont {Lee}},
  \bibinfo {author} {\bibfnamefont {R.~W.}\ \bibnamefont {Erwin}}, \bibinfo
  {author} {\bibfnamefont {S.-H.}\ \bibnamefont {Lee}}, \bibinfo {author}
  {\bibfnamefont {S.}~\bibnamefont {Wakimoto}}, \bibinfo {author}
  {\bibfnamefont {K.~J.}\ \bibnamefont {Thomas}}, \bibinfo {author}
  {\bibfnamefont {M.~A.}\ \bibnamefont {Kastner}}, \ and\ \bibinfo {author}
  {\bibfnamefont {R.~J.}\ \bibnamefont {Birgeneau}},\ }\href {\doibase
  10.1103/PhysRevB.66.014528} {\bibfield  {journal} {\bibinfo  {journal} {Phys.
  Rev. B}\ }\textbf {\bibinfo {volume} {66}},\ \bibinfo {pages} {014528}
  (\bibinfo {year} {2002})}\BibitemShut {NoStop}%
\bibitem [{\citenamefont {Sachdev}(2003)}]{sac03}%
  \BibitemOpen
  \bibfield  {author} {\bibinfo {author} {\bibfnamefont {S.}~\bibnamefont
  {Sachdev}},\ }\href {\doibase 10.1103/RevModPhys.75.913} {\bibfield
  {journal} {\bibinfo  {journal} {Rev. Mod. Phys.}\ }\textbf {\bibinfo {volume}
  {75}},\ \bibinfo {pages} {913} (\bibinfo {year} {2003})}\BibitemShut
  {NoStop}%
\bibitem [{\citenamefont {Caroli}\ \emph {et~al.}(1964)\citenamefont {Caroli},
  \citenamefont {Gennes},\ and\ \citenamefont {Matricon}}]{car64}%
  \BibitemOpen
  \bibfield  {author} {\bibinfo {author} {\bibfnamefont {C.}~\bibnamefont
  {Caroli}}, \bibinfo {author} {\bibfnamefont {P.~D.}\ \bibnamefont {Gennes}},
  \ and\ \bibinfo {author} {\bibfnamefont {J.}~\bibnamefont {Matricon}},\
  }\href {\doibase 10.1016/0031-9163(64)90375-0} {\bibfield  {journal}
  {\bibinfo  {journal} {Physics Letters}\ }\textbf {\bibinfo {volume} {9}},\
  \bibinfo {pages} {307 } (\bibinfo {year} {1964})}\BibitemShut {NoStop}%
\bibitem [{\citenamefont {Shore}\ \emph {et~al.}(1989)\citenamefont {Shore},
  \citenamefont {Huang}, \citenamefont {Dorsey},\ and\ \citenamefont
  {Sethna}}]{sho89}%
  \BibitemOpen
  \bibfield  {author} {\bibinfo {author} {\bibfnamefont {J.~D.}\ \bibnamefont
  {Shore}}, \bibinfo {author} {\bibfnamefont {M.}~\bibnamefont {Huang}},
  \bibinfo {author} {\bibfnamefont {A.~T.}\ \bibnamefont {Dorsey}}, \ and\
  \bibinfo {author} {\bibfnamefont {J.~P.}\ \bibnamefont {Sethna}},\
  }\href@noop {} {\bibfield  {journal} {\bibinfo  {journal} {Phys. Rev. Lett.}\
  }\textbf {\bibinfo {volume} {62}},\ \bibinfo {pages} {3089} (\bibinfo {year}
  {1989})}\BibitemShut {NoStop}%
\bibitem [{\citenamefont {Nakai}\ \emph {et~al.}(2008)\citenamefont {Nakai},
  \citenamefont {Hayashi}, \citenamefont {Ishida}, \citenamefont {Sugawara},
  \citenamefont {Kikuchi},\ and\ \citenamefont {Sato}}]{nak08}%
  \BibitemOpen
  \bibfield  {author} {\bibinfo {author} {\bibfnamefont {Y.}~\bibnamefont
  {Nakai}}, \bibinfo {author} {\bibfnamefont {Y.}~\bibnamefont {Hayashi}},
  \bibinfo {author} {\bibfnamefont {K.}~\bibnamefont {Ishida}}, \bibinfo
  {author} {\bibfnamefont {H.}~\bibnamefont {Sugawara}}, \bibinfo {author}
  {\bibfnamefont {D.}~\bibnamefont {Kikuchi}}, \ and\ \bibinfo {author}
  {\bibfnamefont {H.}~\bibnamefont {Sato}},\ }\href@noop {} {\bibfield
  {journal} {\bibinfo  {journal} {Physica B: Condensed Matter}\ }\textbf
  {\bibinfo {volume} {403}},\ \bibinfo {pages} {1109 } (\bibinfo {year}
  {2008})}\BibitemShut {NoStop}%
\bibitem [{\citenamefont {Kamihara}\ \emph {et~al.}(2008)\citenamefont
  {Kamihara}, \citenamefont {Watanabe}, \citenamefont {Hirano},\ and\
  \citenamefont {Hosono}}]{kam08}%
  \BibitemOpen
  \bibfield  {author} {\bibinfo {author} {\bibfnamefont {Y.}~\bibnamefont
  {Kamihara}}, \bibinfo {author} {\bibfnamefont {T.}~\bibnamefont {Watanabe}},
  \bibinfo {author} {\bibfnamefont {M.}~\bibnamefont {Hirano}}, \ and\ \bibinfo
  {author} {\bibfnamefont {H.}~\bibnamefont {Hosono}},\ }\href@noop {}
  {\bibfield  {journal} {\bibinfo  {journal} {Journal of the American Chemical
  Society}\ }\textbf {\bibinfo {volume} {130}},\ \bibinfo {pages} {3296}
  (\bibinfo {year} {2008})}\BibitemShut {NoStop}%
\bibitem [{\citenamefont {Mazin}\ \emph {et~al.}(2008)\citenamefont {Mazin},
  \citenamefont {Singh}, \citenamefont {Johannes},\ and\ \citenamefont
  {Du}}]{maz08}%
  \BibitemOpen
  \bibfield  {author} {\bibinfo {author} {\bibfnamefont {I.~I.}\ \bibnamefont
  {Mazin}}, \bibinfo {author} {\bibfnamefont {D.~J.}\ \bibnamefont {Singh}},
  \bibinfo {author} {\bibfnamefont {M.~D.}\ \bibnamefont {Johannes}}, \ and\
  \bibinfo {author} {\bibfnamefont {M.~H.}\ \bibnamefont {Du}},\ }\href
  {\doibase 10.1103/PhysRevLett.101.057003} {\bibfield  {journal} {\bibinfo
  {journal} {Phys. Rev. Lett.}\ }\textbf {\bibinfo {volume} {101}},\ \bibinfo
  {pages} {057003} (\bibinfo {year} {2008})}\BibitemShut {NoStop}%
\bibitem [{\citenamefont {Bang}(2010)}]{ban10}%
  \BibitemOpen
  \bibfield  {author} {\bibinfo {author} {\bibfnamefont {Y.}~\bibnamefont
  {Bang}},\ }\href {\doibase 10.1103/PhysRevLett.104.217001} {\bibfield
  {journal} {\bibinfo  {journal} {Phys. Rev. Lett.}\ }\textbf {\bibinfo
  {volume} {104}},\ \bibinfo {pages} {217001} (\bibinfo {year}
  {2010})}\BibitemShut {NoStop}%
\bibitem [{\citenamefont {Bang}(2011)}]{ban11}%
  \BibitemOpen
  \bibfield  {author} {\bibinfo {author} {\bibfnamefont {Y.}~\bibnamefont
  {Bang}},\ }\href@noop {} {\bibfield  {journal} {\bibinfo  {journal}
  {{arXiv}:1112.0142}\ } (\bibinfo {year} {2011})}\BibitemShut {NoStop}%
\bibitem [{\citenamefont {Volovik}(1988)}]{vol88}%
  \BibitemOpen
  \bibfield  {author} {\bibinfo {author} {\bibfnamefont {G.~E.}\ \bibnamefont
  {Volovik}},\ }\href@noop {} {\bibfield  {journal} {\bibinfo  {journal}
  {Journal of Physics C: Solid State Physics}\ }\textbf {\bibinfo {volume}
  {21}},\ \bibinfo {pages} {L221} (\bibinfo {year} {1988})}\BibitemShut
  {NoStop}%
\bibitem [{\citenamefont {Silbernagel}\ \emph {et~al.}(1966)\citenamefont
  {Silbernagel}, \citenamefont {Weger},\ and\ \citenamefont {Wernick}}]{sil66}%
  \BibitemOpen
  \bibfield  {author} {\bibinfo {author} {\bibfnamefont {B.~G.}\ \bibnamefont
  {Silbernagel}}, \bibinfo {author} {\bibfnamefont {M.}~\bibnamefont {Weger}},
  \ and\ \bibinfo {author} {\bibfnamefont {J.~E.}\ \bibnamefont {Wernick}},\
  }\href@noop {} {\bibfield  {journal} {\bibinfo  {journal} {Phys. Rev. Lett.}\
  }\textbf {\bibinfo {volume} {17}},\ \bibinfo {pages} {384} (\bibinfo {year}
  {1966})}\BibitemShut {NoStop}%
\bibitem [{\citenamefont {Silbernagel}\ \emph {et~al.}(1967)\citenamefont
  {Silbernagel}, \citenamefont {Weger}, \citenamefont {Clark},\ and\
  \citenamefont {Wernick}}]{sil67}%
  \BibitemOpen
  \bibfield  {author} {\bibinfo {author} {\bibfnamefont {B.~G.}\ \bibnamefont
  {Silbernagel}}, \bibinfo {author} {\bibfnamefont {M.}~\bibnamefont {Weger}},
  \bibinfo {author} {\bibfnamefont {W.~G.}\ \bibnamefont {Clark}}, \ and\
  \bibinfo {author} {\bibfnamefont {J.~H.}\ \bibnamefont {Wernick}},\
  }\href@noop {} {\bibfield  {journal} {\bibinfo  {journal} {Phys. Rev.}\
  }\textbf {\bibinfo {volume} {153}},\ \bibinfo {pages} {535} (\bibinfo {year}
  {1967})}\BibitemShut {NoStop}%
\bibitem [{\citenamefont {Genack}\ and\ \citenamefont
  {Redfield}(1973)}]{gen73}%
  \BibitemOpen
  \bibfield  {author} {\bibinfo {author} {\bibfnamefont {A.~Z.}\ \bibnamefont
  {Genack}}\ and\ \bibinfo {author} {\bibfnamefont {A.~G.}\ \bibnamefont
  {Redfield}},\ }\href@noop {} {\bibfield  {journal} {\bibinfo  {journal}
  {Phys. Rev. Lett.}\ }\textbf {\bibinfo {volume} {31}},\ \bibinfo {pages}
  {1204} (\bibinfo {year} {1973})}\BibitemShut {NoStop}%
\bibitem [{\citenamefont {Genack}\ and\ \citenamefont
  {Redfield}(1975)}]{gen75}%
  \BibitemOpen
  \bibfield  {author} {\bibinfo {author} {\bibfnamefont {A.~Z.}\ \bibnamefont
  {Genack}}\ and\ \bibinfo {author} {\bibfnamefont {A.~G.}\ \bibnamefont
  {Redfield}},\ }\href@noop {} {\bibfield  {journal} {\bibinfo  {journal}
  {Phys. Rev. B}\ }\textbf {\bibinfo {volume} {12}},\ \bibinfo {pages} {78}
  (\bibinfo {year} {1975})}\BibitemShut {NoStop}%
\bibitem [{\citenamefont {Wortis}(1998)}]{wor98}%
  \BibitemOpen
  \bibfield  {author} {\bibinfo {author} {\bibfnamefont {R.}~\bibnamefont
  {Wortis}},\ }\href@noop {} {Ph.D. thesis},\ \bibinfo  {school} {University of
  Illinois Champaign Urbana} (\bibinfo {year} {1998})\BibitemShut {NoStop}%
\end{thebibliography}%
\end{document}